       \documentclass[twocolumn,prb,aps,a4paper,epsf,showkeys,superscriptaddress] {revtex4} 

\usepackage{etoolbox}
\usepackage{fancyhdr} 
\newtoggle{usualpreprint} \newtoggle{journal-like} \newtoggle{journal-like-bigfigures}

   \togglefalse{usualpreprint} \togglefalse{journal-like} \toggletrue{journal-like-bigfigures}  

\newcommand{\myheading}{IVCT luminescence: Interplay between anomalous and $5d$--$4f$ emissions in Yb-doped fluorites}
\newcommand{\authors}{Barandiar\'an and Seijo}
  \newcommand{\FIGa}  {ivct-gs-Qet}
  \newcommand{\FIGb}  {monomers-dimer}
  \newcommand{\FIGc}  {CaF2-monomers-and-YbIIYbIII-11}
  \newcommand{\FIGd}  {CaF2-YbIIYbIII-11-few}
  \newcommand{\FIGe}  {CaSrF2-IVCTL}
  \newcommand{\FIGf}  {ivct-scheme}
  \newcommand{\FIGg}  {CaSrF2-ESA}
  \newcommand{\FIGh}  {CaSrBaF2-YbIIYbIII-11-few}
  \newcommand{\FIGi}  {SrF2Cl2-YbIIYbIII-11-few}
  \newcommand{\FIGj}  {Yb-cases}
\newcommand{\Luiscolor}{magenta}   

\usepackage[final]{graphicx}   
\usepackage{dcolumn}           
\usepackage{bm}                    
\usepackage{longtable} 
\usepackage{rotating}
\usepackage{color}
\usepackage{comment}
\usepackage{epstopdf}
\usepackage[bitstream-charter]{mathdesign}
\usepackage[T1]{fontenc}

\def\CASPT2{ANDERSSON:90,ANDERSSON:92}
\def\MSCASPT2{FINLEY:98,ZAITSEVSKII:95}

\def\abinitio{{\it ab initio}}

\def\cmm1{~cm$^{-1}$}
\def\etal{{\em et al.}}

\def\ec{embedded-cluster}

\def\nua1g{$\bar{\nu}_{a_{1g}}$}
\def\nue{$\bar{\nu}_{a_{1g}}$}

\def\Te{$T_e$}
\def\DRet2g{$\Delta R_e(t_{2g}-f)$}

\def\Dth{$D_{{\rm 2}h}$}

\def\Oh{$O_h$}

\def\Gsixu{$\Gamma_{6u}$}
\def\Gsevu{$\Gamma_{7u}$}
\def\Geigu{$\Gamma_{8u}$}

\def\sAog{$^1A_{\rm 1g}$}

\def\Aog{$A_{\rm 1g}$}

\def\Aou{$A_{\rm 1u}$}
\def\Atu{$A_{\rm 2u}$}

\def\Eu{$E_{\rm u}$}

\def\Tou{$T_{\rm 1u}$}

\def\Ttu{$T_{\rm 2u}$}

\def\dF{${^2F}$}

\def\dAtu{${^2A}_{2u}$}
\def\dTou{${^2T}_{1u}$}
\def\dTtu{${^2T}_{2u}$}

\def\Batp{Ba$^{2+}$}
\def\Catp{Ca$^{2+}$}

\def\Eutp{Eu$^{2+}$}

\def\Mtp{M$^{2+}$}

\def\Srtp{Sr$^{2+}$}

\def\Tmtp{Tm$^{2+}$}

\def\Ybtp{Yb$^{2+}$}

\def\Cethp{Ce$^{3+}$}

\def\Euthp{Eu$^{3+}$}

\def\Prthp{Pr$^{3+}$}

\def\Ybthp{Yb$^{3+}$}

\def\Cefp{Ce$^{4+}$}

\def\Prfp{Pr$^{4+}$}

\def\Ufp{U$^{4+}$}

\def\Lnnp{Ln$^{n+}$}
\def\Lnnpop{Ln$^{(n+1)+}$}

\def\Fm{F$^-$}

\def\CstNaYCl6{Cs$_2$NaYCl$_6$}
\def\CstNaLnCl6{Cs$_2$NaLnCl$_6$}
\def\CstNaYBr6{Cs$_2$NaYBr$_6$}
\def\KtNaGaF6{K$_2$NaGaF$_6$}
\def\KtNaScF6{K$_2$NaScF$_6$}
 
\def\CsCaBr3{CsCaBr$_3$}
\def\CsCaF3{CsCaF$_3$}
\def\KCdF3{KCdF$_3$}
\def\KMgF3{KMgF$_3$}
\def\KZnF3{KZnF$_3$}

\def\KtCrO4{K$_2$CrO$_4$}
\def\KtFeO4{K$_2$FeO$_4$}
\def\KtSO4{K$_2$SO$_4$}
\def\KtSeO4{K$_2$SeO$_4$}

\def\CstGeF6{Cs$_2$GeF$_6$}
\def\CstZrCl6{Cs$_2$ZrCl$_6$}
\def\CstUCl6{Cs$_2$UCl$_6$}
\def\CstUBr6{Cs$_2$UBr$_6$}
\def\CstZrBr6{Cs$_2$ZrBr$_6$}

\def\CstUOtCl4{Cs$_2$UO$_2$Cl$_4$}
\def\TMAtUCl6{(TMA)$_2$UCl$_6$}
\def\YF3{YF$_3$}

\def\abinitio{{\it ab initio}}
\def\ec{embedded-cluster}

\def\etal{{\it et al.}}

\def\cmm1{cm$^{-1}$}

\def\nue{$\omega_{a_{1g}}$}
\def\Te{T$_{e}$}

\def\YbII{Yb$^{2+}$}
\def\YbIII{Yb$^{3+}$}

\def\Fm{F$^{-}$}

\def\OCeiv+xiv{(OCe$_4$)$^{14+}$}
\def\OUiv+xiv{(OU$_4$)$^{14+}$}

\def\Ybivp{\mbox{\Ybtp--\Ybthp}}
\def\Ybiipiiip{\mbox{\Ybtp--\Ybthp}}
\def\Ybiiipiip{\mbox{\Ybthp--\Ybtp}}

\def\Dclus{(YbX$_8$)$^{6-}$}
\def\DclusL{(YbX$_8$)$^{6-}_L$}
\def\DclusR{(YbX$_8$)$^{6-}_R$}
\def\Aclus{(YbX$_8$)$^{5-}$}
\def\AclusL{(YbX$_8$)$^{5-}_L$}
\def\AclusR{(YbX$_8$)$^{5-}_R$}
\def\Qet{$Q_{et}$}
\def\Qetzz{$Q_{et}^{00}$}
\def\Qetij{$Q_{et}^{ij}$}
\def\dFsh{$^2F_{7/2}$}
\def\dFfh{$^2F_{5/2}$}
\def\dfem{$5d$--$4f$}
\def\fdabs{$4f$--$5d$}

\def\dL{$d_L$}
\def\dR{$d_R$}

\def\dYbX{$d_{\rm Yb-X}$}

\iftoggle{journal-like}           { \pagestyle{fancy}  \lhead{\authors}\chead{}\rhead{\myheading}\lfoot{}\cfoot{\thepage}\rfoot{} } {}
\iftoggle{journal-like-bigfigures}{ \pagestyle{fancy}  \lhead{\authors}\chead{}\rhead{\myheading}\lfoot{}\cfoot{\thepage}\rfoot{} } {}

\iftoggle{journal-like}           { \def\mylongtablefootnote{0.92\textwidth} } {}
\iftoggle{journal-like-bigfigures}{ \def\mylongtablefootnote{0.92\textwidth} } {}
\iftoggle{usualpreprint}          { \def\mylongtablefootnote{0.96\textwidth} } {}

\begin{document}
\title{ 
       Intervalence Charge Transfer Luminescence: Interplay between anomalous 
       and $5d-4f$ emissions in Yb-doped 
       fluorite-type crystals.
      }
\date{\today}
 \author{Zoila Barandiar\'an}
 \thanks{Corresponding author}
 \email{zoila.barandiaran@uam.es}
 \author{Luis Seijo}
 \affiliation{Departamento de Qu\'{\i}mica,
              Universidad Aut\'onoma de Madrid, 28049 Madrid, Spain}
 \affiliation{Instituto Universitario de Ciencia de Materiales Nicol\'as Cabrera,
              Universidad Aut\'onoma de Madrid, 28049 Madrid, Spain}
   \keywords{{\em Ab initio}, IVCT, electron transfer, Yb$^{2+}$, Yb$^{3+}$, fluorites, anomalous emission, luminescence
            }
\begin{abstract}
In this paper we report the existence of intervalence charge transfer (IVCT)
luminescence in Yb-doped fluorite-type crystals associated with \Ybiipiiip\
mixed valence pairs. 
  By means of embedded cluster, wave function theory \abinitio\ calculations, we 
show that the widely studied, very broad band,
anomalous emission of \Ybtp-doped CaF$_2$ and SrF$_2$, usually associated with
impurity-trapped excitons,
is, rather, an IVCT luminescence associated with \Ybiipiiip\ mixed valence pairs.
The IVCT luminescence is 
very efficiently excited by a two-photon upconversion mechanism where each
photon provokes the same strong $4f^{14}$--1\Aog$\rightarrow$$4f^{13}($\dFsh$)5de_g$--1\Tou\ 
absorption in the \Ybtp\ part of the pair: the first one, from
the pair ground state;
the second one, from an excited state of the pair whose \Ybthp\ moiety is in the higher $4f^{13}($\dFfh$)$
multiplet.
The  
  \Ybiipiiip\ $\rightarrow$ \Ybiiipiip\ 
IVCT emission consists of an
\mbox{\Ybtp\ $5de_g$ $\rightarrow$ \Ybthp\ $4f_{7/2}$} charge transfer accompanied by 
  a $4f_{7/2} \rightarrow 4f_{5/2}$ deexcitation within the \Ybtp\ $4f^{13}$ subshell:
  [\dFfh$5de_g$,\dFsh] $\rightarrow$ [\dFsh,$4f^{14}$].
The 
IVCT vertical transition leaves the 
oxidized and reduced 
moieties of the pair after electron transfer
very far from their equilibrium structures; 
this explains the unexpectedly large band width of the emission band and its
low peak energy, because the large reorganization energies are substracted from the normal emission.
  The IVCT energy diagrams resulting from the quantum mechanical calculations explain the different luminescent properties
  of Yb-doped CaF$_2$, SrF$_2$, BaF$_2$, and SrCl$_2$:
  the presence of IVCT luminescence in Yb-doped CaF$_2$ and SrF$_2$;
  its coexistence with regular $5d$-$4f$ emission in SrF$_2$;
  its absence in BaF$_2$ and SrCl$_2$;
  the quenching of all emissions in BaF$_2$;
  and the presence of additional \dfem\ emissions in SrCl$_2$ which are absent in SrF$_2$.
They also allow to interpret and reproduce recent experiments on transient photoluminescence enhancement
in \Ybtp-doped CaF$_2$ and SrF$_2$, the appearence of \Ybtp\ \fdabs\ absorption
bands in the excitation spectra of the IR \Ybthp\ emission in partly reduced
CaF$_2$:\Ybthp\ samples, and to identify the broad  band observed in
the excitation spectrum of the so far called anomalous emission of SrF$_2$:\Ybtp\
as an IVCT absorption, which corresponds to an
\Ybtp\ $4f_{5/2}$ $\rightarrow$ \Ybthp\ $4f_{7/2}$ electron transfer.

\end{abstract}
\maketitle
\renewcommand{\thefootnote}{\alph{footnote}}
\section{\label{SEC:intro}Introduction}
The capability of lanthanide ion dopants to luminesce
from their $4f^{N}$ and/or  $4f^{N-1}5d$ excited states has granted
them a prominent role as activators in solid state lighting,
lasers, fiber  amplifiers, and medical imaging devices.~\cite{EIJK:02,LIU:05,RONDA:07,WEBER:02}
Frequently, however, their applicability is compromised by
quenching or by replacement of the expected luminescence
by an anomalous emission.
Detailed understanding of these mechanisms is important
in the search for new interesting phosphor and scintillating
materials based on lanthanide ion activators.

The anomalous emission of \Ybtp\ in Yb-doped materials is a prototypical case
(for a review on the anomalous emission of \Ybtp\ and \Eutp-doped crystals, see Ref.~\onlinecite{DORENBOS:03:a}).
%
%
The interplay between anomalous and \dfem\ emissions in \Ybtp-doped 
fluorite-type crystals results in a complex electronic spectroscopy,
which has been the focus of investigations for 
decades.~\cite{FEOFILOV:56,KAPLYANSKII:62,REUT:76,MCCLURE:85,MOINE:88,MOINE:89,DORENBOS:03:a,KACZMAREK:05,REID:11,SENANAYAKE:13}
The adjective anomalous has been used to reflect the irregular, unexpected, very large
Stokes shift of the emission observed,
with respect to the $4f^{14}\rightarrow 4f^{13}5d$ excitation, which makes
the red shifted emission band extremely broad. 
When \Ybtp\ is doped in CaF$_2$, the anomalous emission 
prevails;~\cite{FEOFILOV:56,KAPLYANSKII:62,REUT:76,MOINE:89,REID:11}
it also occurs in SrF$_2$, where
both, anomalous emission and regular
$4f^{13}5d\rightarrow 4f^{14}$ luminescence have been detected 
together;~\cite{REUT:76,MCCLURE:85,MOINE:88,SENANAYAKE:13}
and no emission whatsoever occurs in 
BaF$_2$ up to 1.5$\mu$m.~\cite{REUT:76,MOINE:89} 
Changing the ligand, as in the 
SrF$_2$ and SrCl$_2$ series, also affects the
interplay.
The dual character of the luminescence of Yb in SrF$_2$  
disappears
in the SrCl$_2$ host, where the anomalous emission is not observed; only regular
\dfem\ emission bands have been assigned in this  case showing a complex temperature
behaviour.~\cite{WITZKE:73,PAN:08} 
Analyses of spectroscopic data have suggested that the anomalous
emission observed in \Ybtp-doped CaF$_2$ and SrF$_2$ 
is associated with an excited state which 
has a "radically different" (smaller)
radiative rate compared with that of higher lying levels.~\cite{MOINE:89,REID:11} 
This characteristic has allowed the application
of a two-frequency UV + IR transient photoluminescence enhancement technique
which ultimately produces the IR excited state absorption spectrum (ESA)
from the lowest anomalous state to close lying energy levels.~\cite{REID:11,SENANAYAKE:13}
The anomalous emission of Yb has also been observed 
in \Ybthp-doped CaF$_2$ after application of 
reducing conditions.~\cite{KACZMAREK:05}

Analogous competitions between regular \dfem\ (as well as $4f$--$4f$) and anomalous emissions
have been reported involving other commonly used lanthanide ion activators:
\Cethp, \Prthp, and \Eutp.~\cite{REUT:76,MCCLURE:85,DORENBOS:03:a,BESSIERE:04,BESSIERE:06,GRINBERG:11}
For all of them (including \Ybtp), it is recognized that 
it is difficult to prevent the coexistence of another
different valence state,~\cite{LOH:68,LOH:69,WITZKE:73,REUT:76,PAN:08,SU:05:a,SU:05:b}
which means that 
the formation of mixed valence pairs, that is,  
pairs between two metal sites differing only in
oxidation state, is very likely.
In this way, occurrence of \Cethp--\Cefp, \Prthp--\Prfp, \Eutp--\Euthp,  and
\Ybtp--\Ybthp\ mixed valence pairs should be expected, to some extent.
In this paper we show that the study of the electronic structure of 
 \Ybtp--\Ybthp\ mixed valence pairs is crucial to understand
anomalous emission and its interplay
with regular emissions in Yb-doped fluorite-type crystals. The same is true for \Cethp--\Cefp\ pairs,
as we show in an analogous study of the anomalous emission in
\Cethp-doped Cs$_2$LiLuF$_6$ elpasolite.~\cite{SEIJO:UP}

Since mixed valence pairs (for lanthanides:
\mbox{\Lnnp-\Lnnpop}) constitute a donor--acceptor ($DA$)
redox system, electron transfer between the donor and
acceptor sites,
Ln$^{n+}$ + Ln$^{(n+1)+}$ $\rightarrow$ Ln$^{(n+1)+}$ + Ln$^{n+}$,
may occur, and the process can be referred to as an intervalence 
charge transfer (IVCT),~\cite{VERHOEVEN:96} in analogy with the IVCT
processes that have been investigated
involving mostly mixed valence transition metal 
compounds~\cite{MARCUS:64,ALLEN:67,HUSH:67,ROBIN:68,PIEPHO:78}
and, scarcely, lanthanide ions.~\cite{BLASSE:91,VANSCHAIK:93}
Although most IVCT studies refer to thermally and radiatively
induced electron transfer between the ground states of the
donor and acceptor sites, we show here that the $D + A \rightarrow A + D$ reaction 
can involve many of the excited electronic states of the mixed valence pair
and lead to a variety of non-radiative and radiative IVCT processes.

The goal of this paper is to show that all of the spectral features we
have summarized above for \mbox{\Ybtp-doped} CaF$_2$,  SrF$_2$,  BaF$_2$, and  SrCl$_2$
hosts can be explained if 
the existence of \Ybtp--\Ybthp\ mixed valence pairs is assumed, \abinitio\
multielectronic wavefunction-based diabatic potential energy surfaces of the ground and excited states of the embedded
\Ybtp--\Ybthp\ active pairs
are calculated, and 
quantitative energy diagrams for the 
\mbox{Yb$^{2+}$ + Yb$^{3+}$ $\rightarrow$ Yb$^{3+}$ + Yb$^{2+}$}
electron transfer reaction, along
well defined reaction coordinates, are produced. 
In this framework, the anomalous emission of Yb-doped fluorite-type crystals 
is interpreted as 
  an \Ybiipiiip\ $\rightarrow$ \Ybiiipiip\ IVCT emission, in which
an \mbox{\Ybtp\ $5de_g$ $\rightarrow$ \Ybthp\ $4f_{7/2}$} electron transfer and 
  a $4f_{7/2} \rightarrow 4f_{5/2}$ deexcitation within the \Ybtp\ $4f^{13}$ subshell take place:
  [\dFfh$5de_g$,\dFsh] $\rightarrow$ [\dFsh,$4f^{14}$].
So, the bielectronic character of this IVCT luminescence
explains the very slow radiative decay rate of the so far called anomalous emission.
Its very large band width comes from the large change in the electron transfer
reaction coordinate accompanying the electronic transition: 
$|\Delta Q_{et}|$= 0.49~\AA\ (CaF$_2$), 0.60~\AA\ (SrF$_2$).
Its occurrence and temperature quenching in
CaF$_2$ and SrF$_2$, including its coexistence with regular
\dfem\ emission in SrF$_2$ below 140~K, or its abscence in BaF$_2$ and SrCl$_2$,
including the absence of any emission in BaF$_2$,  can be
understood on the basis of changes in the topology of the diabatic
IVCT energy diagrams of the \Ybtp--\Ybthp\ invervalence pairs in 
the different hosts. 
The IR ESA spectra measured in CaF$_2$ and SrF$_2$ are also
reproduced, showing a close, upper lying state, which may contribute a much
faster monoelectronic \mbox{\Ybtp\ $5de_g$ $\rightarrow$ \Ybthp\ $4f_{5/2}$}
IVCT emission, plus a number of other close lying upper levels which
contribute to the transient photoluminescence enhancement.

In parallel to this work, we are presenting analogous studies on 
the nature of the anomalous luminescence of
Ce-doped Cs$_2$LiLuCl$_6$ and its interplay with regular emissions.~\cite{SEIJO:UP}
There, the fast anomalous emission of Ce is identified as a monoelectronic
\mbox{\Cethp\ $5de_g$ $\rightarrow$ \Cefp\ $4f$} 
IVCT luminescence, which is observed and calculated to be above
the lowest, regular $5de_g$ $\rightarrow$ $4f$ emission of the
\Cethp\ active center.   
That work and the present paper are part of an effort focused on
assessing and showing the relevance of intervalence charge 
transfer processes in the optical properties of very common
lanthanide ions activators. Both works show that 
IVCT luminescence has been observed experimentally, but has not been identified
as such, neither in \Ybtp-doped nor in \Cethp-doped crystals. 
The same is true for other IVCT processes which are
predicted and have not been identified or observed (i.e. 
transient IVCT photoluminescence enhancement, other IVCT
absorptions, emissions,
and non-radiative energy transfers).
Altogether, the results presented so far suggest that
the theoretical effort should continue and cover
other common hosts and activators like those mentioned above;
they also suggest that incorporating quantitative IVCT energy diagrams
to experimental analyses should be useful.~\cite{BARANDIARAN:UP}

  This paper is organized as follows:
  In Sec.~\ref{SEC:method} we describe the \abinitio\ calculation of 
  intervalence charge transfer potential energy surfaces of \YbII-\YbIII\ pairs in fluorite-type crystals
  and the corresponding IVCT configuration coordinate diagrams.
  In Sec.~\ref{SEC:IVCT-CaF2} we interpret the anomalous luminescence of CaF$_2$:\Ybtp\ as an IVCT luminescence,
  we explain its mechanism and we discuss the involvement of IVCT states 
  in transient photoluminescence enhancement experiments and
  in the excitation of the IR \Ybthp\ emission by \Ybtp\ \fdabs\ absorptions.
  Finally, in Sec.~\ref{SEC:interplay} we show that changes in the topology of the IVCT energy diagrams
  account for the interplay between IVCT and regular \dfem\ emissions in the chemical series of fluorite-type crystals 
  CaF$_2$, SrF$_2$, BaF$_2$ (Sec.~\ref{SEC:CaSrBaF2}) and SrF$_2$, SrCl$_2$ (Sec.~\ref{SEC:SrF2SrCl2}).
  The conclusions are summarized in Sec.~\ref{SEC:Conclusions}.

\section{Potential energy surfaces of \Ybiipiiip\ active pairs in fluorite-type crystals}
\label{SEC:method}

In this Section we describe how the diabatic potential energy surfaces of
the ground and excited states of 
\Ybiipiiip\ mixed valence active pairs can be calculated in the 
CaF$_2$, SrF$_2$, BaF$_2$, and SrCl$_2$ (MX$_2$) hosts using the 
potential energy surfaces obtained from independent  \Ybtp\ 
and \Ybthp\ embedded cluster calculations as building blocks.
From them, quantitative IVCT energy diagrams can be built in terms
of normal electron transfer reaction coordinates which involve
concerted vibrational breathing modes of the donor \Dclus\ and
acceptor \Aclus\ sites.
The approximations involved in this
procedure are outlined; their accuracy is expected
to be sufficient to achieve the goals of this
work, which have been stated in the Introduction.

\subsection{\Ybiipiiip\ diabatic potential energy surfaces}
\label{SEC:diabatic_PES}

A convenient definition for the diabatic wavefunctions and energies of 
the \Ybivp\ mixed valence pairs can be set up using 
generalized antisymmetric product functions~\cite{MCWEENY:59}
resulting from the combination of the $n_D$ states of the donor $D$,
and $n_A$ states of the acceptor $A$ sites, which in this work are the 
states of the 
$4f^{14}$,
$4f^{13}5de_g$,
$4f^{13}5dt_{2g}$, and
$4f^{13}a_{1g}^{ITE}$ manifolds of the donor
\Dclus\ 
and those of the
$4f^{13}$ manifold of the acceptor
\Aclus\ separated embedded clusters, respectively.
So, from  the combination of the state $i$ of $D$, $\Phi_{Di}$, 
and the state $j$ of $A$, $\Phi_{Aj}$,
two diabatic wavefunctions are obtained:
    one for the $ij$ state of $DA$, $M \hat A (\Phi_{Di} \Phi_{Aj})$,
and one for the $ji$ state of $AD$, $M \hat A (\Phi_{Aj} \Phi_{Di})$ 
($M$ is a normalization constant and $\hat A$ is the inter-group antisymmetrization operator~\cite{MCWEENY:59}),
where $DA$ and $AD$ refer to the mixed valence pairs before
and after electron transfer: \Ybivp\ and \mbox{\Ybthp--\Ybtp}, respectively.
These two diabatic wavefunctions will be needed
to study the electron transfer 
\mbox{\DclusL--\AclusR$\rightarrow$\AclusL--\DclusR} from the left to the
right moieties of the mixed valence pair embedded in the MX$_2$ solid.
Altogether, a basis of 2$n_Dn_A$ diabatic wavefunctions are obtained.
The   two corresponding diabatic potential energy surfaces 
$E_{DiAj}^{\rm diab}$ and  $E_{AjDi}^{\rm diab}$
are the 
expected values of the fixed nuclei Hamiltonian of the embedded pair $\hat{H}$:
\mbox{$E_{DiAj}^{\rm diab} = \langle M\hat A (\Phi_{Di} \Phi_{Aj}) | \hat H | M\hat A (\Phi_{Di} \Phi_{Aj}) \rangle$,}
\mbox{$E_{AjDi}^{\rm diab} = \langle M\hat A (\Phi_{Aj} \Phi_{Di}) | \hat H | M\hat A (\Phi_{Aj} \Phi_{Di}) \rangle$.}
Given that
the electronic spectroscopic transitions are dominated by the totally symmetric vibrational coordinates,
the breathing modes of the donor \Dclus\ and acceptor \Aclus\ sites are the only
vibrational degrees of freedom we will consider, which results in the parametric
dependence of the diabatic potential energy surfaces on the Yb--X distance in
the left and right components of the $DA$ and $AD$ pairs,
$d_L$ and $d_R$:
$E_{DiAj}^{\rm diab}(d_L,d_R)$ and  $E_{AjDi}^{\rm diab}(d_L,d_R)$.
Note that $E_{DiAj}^{\rm diab}(x,y) = E_{AjDi}^{\rm diab}(y,x)$.
The two diabatic potential energy surfaces for the ground state $00$ of the
\Ybiipiiip\ mixed valence pair in Yb-doped CaF$_2$,
$E_{D0A0}^{\rm diab}(d_L,d_R)$ and  $E_{A0D0}^{\rm diab}(d_L,d_R)$,
can be seen in Fig.~\ref{FIG:ivct-gs-Qet}.  

It is important to note that in this work
the electronic coupling between any two diabatic states 
$ij$ of $DA$ and $kl$ of $AD$,
\mbox{$V_{DiAj,AkDl}^{\rm diab} = \langle M\hat A (\Phi_{Di} \Phi_{Aj}) | \hat H | M\hat A (\Phi_{Ak} \Phi_{Dl}) \rangle$,}
will be neglected for all values of the $d_L$ and $d_R$ coordinates.
As a consequence,
there will be no avoided crossings among the
diabatic potential energy surfaces nor mixings between the diatabic states.
Whereas this approximation can be expected to be a good one away
from the intersection regions, 
it becomes weaker on them, since 
the diabatic energies at intersections give upperbounds
to the energy barriers that would result from avoided crossings should the electronic
couplings be considered.
We comment further on this issue in the next section where
diabatic energy diagrams for the electron transfer reactions will be
extracted from the diabatic energy surfaces.
As we will show in this work, many spectroscopic features of
the mixed valence pairs can be addressed quantitatively or semiquantitatively
within the diabatic approximation, using the diabatic energy
surfaces only. We describe next how they will be computed in this work.

According to the group function theory used to define the diabatic wavefunctions
of the mixed valence pairs,
the diabatic pair energies are the sum of the donor and acceptor energies 
plus their mutual Coulomb and exchange interaction.~\cite{MCWEENY:59}
The latter should be almost independent of the donor and acceptor states, in general.
Hence, we can write:
\begin{eqnarray}
E_{DiAj}^{\rm diab} = E_{Di} + E_{Aj} + E^{cx}_{DiAj} \approx E_{Di} + E_{Aj} + E^{cx}_{DA}
\,.
\label{EQ:EDA}
\end{eqnarray} 
In Eq.~\ref{EQ:EDA}, $E_{Di}$ and $E_{Aj}$ include the embedding interactions of $D$ and $A$ with the crystalline environment 
of the $DA$ pair in the MX$_2$ crystal.
In this work, we will compute $E_{DiAj}^{\rm diab}$ 
using the energy curves obtained from independent donor \Dclus\
and acceptor \Aclus\ clusters embedded in the perfect
MX$_2$ hosts, as building blocks.
According to this alternative we will use:
\begin{eqnarray}
E_{DiAj}^{\rm diab}(d_L,d_R) = E_{Di}^{(iec)}(d_L) + E_{Aj}^{(iec)}(d_R) + E^{(iec)}_0(d_{DA})
\,,
\nonumber\\
E_{AjDi}^{\rm diab}(d_L,d_R) = E_{Aj}^{(iec)}(d_L) + E_{Di}^{(iec)}(d_R) + E^{(iec)}_0(d_{DA})
\,,
\label{EQ:E02}
\end{eqnarray}
where $E_{Di}^{(iec)}(d_L)$ and $E_{Aj}^{(iec)}(d_L)$ are the independent embedded cluster
energy curves of the donor and acceptor obtained as described in Sec.~\ref{SEC:iecmeth}, and
\begin{eqnarray}
E^{(iec)}_0(d_{DA}) &=& E^{cx}_{DA}(d_{DA}) - E^{cx}_{DM}(d_{DA}) - E^{cx}_{AM}(d_{DA})
\nonumber\\
&\approx& (q_D \times q_A - (q_D+q_A) \times q_M) e^2/d_{DA}
\,,
\end{eqnarray}
with
$E^{cx}_{DM}$ (or $E^{cx}_{AM}$) standing for the Coulomb and exchange interaction energy between the 
donor \Dclus\ (or acceptor \Aclus) cluster and
the cluster of 
the original host cation (MX$_8)^{6-}$ (M: \Catp, \Srtp, or \Batp, in this case). 
Except for short cation-cation distances,
$- E^{cx}_{DM} - E^{cx}_{AM} + E^{cx}_{DA}$
$\approx (- q_D \times q_M - q_A \times q_M + q_D \times q_A) e^2/d_{DA}$
$= (- 2 \times 2 - 3 \times 2 + 2 \times 3) e^2/d_{DA}$
$=- (2\times2) e^2/d_{DA}$.
In any case,
the term $E^{(iec)}_0(d_{DA})$ is common to the $DA$ and $AD$ energy surfaces and to all states of both.
Its effect is a common shift of all of them and, consequently, it does not contribute to energy differences between them.

In this alternative, 
the symmetry reductions around $D$ and $A$ due to the presence of the other ($A$ and $D$, respectively)
are not considered.
Therefore, the energy splittings driven by symmetry lowering,
which are dependent on the distance and relative orientation between $D$ and $A$,
are not obtained.
The most important ones in the present work would be the splittings produced
on the \Dclus\ levels by the presence of the \Ybthp\ substitutional ion. 
Yet, whereas such site symmetry reductions
would add significant conputational effort to the embedded cluster calculations,
the corresponding splittings would only be responsible for fine structure features of the
spectra; they would neither affect the positions nor the number of the main 
absorption and emission bands. Therefore, 
the alternative chosen in this work will be able
to capture the basics of the energy surfaces of the pairs by means of
completely independent calculations on the $D$ and $A$ embedded clusters.
This is the strength of the present approach.

\subsection{Independent embedded cluster calculations}
\label{SEC:iecmeth}
In this Section we describe the details and summarize the results 
of the quantum mechanical calculations of the
$E_{Di}$(\dYbX)  and $E_{Aj}$(\dYbX) components of the mixed valence pair energies in
Eq.~\ref{EQ:E02}.
As donor $D$ and acceptor $A$ we adopted, respectively, the \Dclus\ and \Aclus\ clusters
(X= F, Cl) .
We performed \abinitio\ wave function theory embedded cluster quantum chemical calculations
on these  clusters embedded in the MX$_2$ hosts,
with the MOLCAS suite of programs.~\cite{MOLCAS}
The calculations include bonding interactions, static and dynamic electron correlation effects,
and scalar and spin-orbit coupling relativistic effects within the clusters,
which are treated at a high theoretical level.
They also include Coulomb, exchange, and Pauli repulsion interations between the host and the clusters,
which are considered at a lower theoretical level by means of a quantum mechanical embedding potential.
Electron correlation effects between the cluster and the host are excluded from these calculations.

\subsubsection{
\label{SEC:detailsQMC}
Details of the quantum mechanical calculations
}

The calculations presented here assume that \Ybtp\ and \Ybthp\ ions
substitute for \Mtp\ (\Mtp= \Catp, \Srtp, \Batp) and occupy cubic sites in
the perfect fluorite structures.~\cite{WYCKOFF:82}
Following the \ec\ approximation, the imperfect
crystals were divided into the defect cluster and the 
embedding host, which were represented as follows.

The \ec\ scalar relativistic second-order Douglas-Kroll-Hess 
Hamiltonian~\cite{DOUGLAS:74,HESS:86} and 
wave functions, comprising the Yb impurity at (0,0,0)
and eight X ligands at variable (x,x,x), 
included all the electrons of \Ybtp\ or \Ybthp\
and eight X$^-$.
The basis sets used to expand the cluster molecular orbitals
included the all electron ANO-RCC bases
Yb (25s22p15d11f4g2h) [9s8p5d4f3g2h],
F  (14s9p4d3f2g) [5s4p3d]
or Cl (17s12p5d4f2g) [6s5p3d].~\cite{ROOS:05,ROOS:08}
In addition,
the highest occupied $s$ and $p$ orbitals of the
embedding \Mtp\ ions, contracted as
\Catp\ (20s15p)[1s1p],
\Srtp\ (23s19p)[1s1p],
\Batp\ (27s23p)[1s1p],
were used as orthogonalization functions
at the 12 second neighbour sites ($\frac{1}{2}$,$\frac{1}{2}$,0),
to fulfil strong orthogonality,  and
5 $s$-type Gaussian type functions were used at the
six ($\frac{1}{2}$,0,0) interstitial sites surrounding the YbX$_8$ 
cube in the fluorite structures; their
orbital exponents were optimized to give minimal impurity-trapped
exciton energy.
The \ec\ calculations were performed using \Dth\
symmetry.

AIMP embedding potentials were used in the \ec\
Hamiltonian to represent the host effects
due to interactions with the remainder of the crystal 
ions, which include quantum mechanical embedding effects
associated with exchange and Pauli repulsion, in addition to
Coulomb electron repulsion and Madelung interactions.~\cite{BARANDIARAN:88,SEIJO:99}
The embedding potentials for all four MX$_2$ hosts were obtained in this work
%
to represent the \Mtp\ and X$^-$ ions
located outside the cluster, at their cubic crystal  structure
sites [Group 225, $Fm\bar 3m$, $a_0$ = 
5.46294~\AA\ (CaF$_2$),
5.796~\AA\   (SrF$_2$),
6.2001~\AA\  (BaF$_2$),
6.9744~\AA\ (SrCl$_2$)~\cite{WYCKOFF:82}];
the potentials were obtained by performing
self-consistent embedded \Mtp\ and \Fm\ ions calculations at the
Hartree-Fock level on
the perfect host crystal as described in Ref.~\onlinecite{SEIJO:99}.
All ions located in a cube of
7$\times$7$\times$7 unit cells, centered at the impurity site,
were represented by their total ion embedding AIMP;
an additional set of 2781 point charges 
was used to ensure that the Ewald potential is
reproduced within the cluster volume. These charges were obtained
following the zero multipole method of Gell\'e and Le Petit.~\cite{GELLE:08}


A recent study on radial correlation effects on interconfigurational
transitions at the end of the lanthanide series has revealed that
$4f$ to $5f$ double excitations must be included at the variational
multiconfigurational self-consistent field step preceding
second order perturbation theory calculations.~\cite{BARANDIARAN:13}
Consequently, state-average
restricted active space self-consistent field 
(RASSCF)~\cite{OLSEN:88,MALMQVIST:90,MALMQVIST:08}
calculations were done on the \Dclus\ and \Aclus\ embedded clusters
including $4f$ to $5f$ single and double excitations
from the $4f^{14}$ and $4f^{13}$ reference to calculate the $4f^{14}$--\sAog\ ground state, 
and the \dAtu, \dTou, and \dTtu\ states of the $4f^{13}$ manifold,
whereas
single excitations
from the $4f^{14}$ reference to one $a_{1g}$ shell and to
the $5d$ shell (which belong to the $e_g$ and $t_{2g}$
\Oh\ irreducible representations) were also allowed to provide the minimal
configurational space required for the spectroscopy
of the \Dclus\ cluster, plus
additional
single and double excitations from the $4f$ shell
to the $5f$ shell to account for the large change
of radial correlation upon
$4f^{14}\rightarrow 4f^{13}5d^{1}$ electronic transitions.
These calculations are referred here as
RASSCF($4f$-$3h$/$5f$/$a_{1g}$$e_g$$t_{2g}$-$1e$), which indicates
the maximum number of holes allowed in the $4f^{14}$ shell
and the maximum number of electrons allowed in the 
empty $a_{1g}$ and 5d shells to calculate the
$ungerade$ $^{2S+1}\Gamma _u$ states of the $\Gamma$ =
\Aou, \Atu, \Eu, \Tou, \Ttu\ octahedral irreps.

Once these wave functions are obtained, they become the multireference
for multistate second onder perturbation theory
calculations (MS-RASPT2),~\cite{ZAITSEVSKII:95,FINLEY:98,MALMQVIST:08}
which include dynamic electron correlation, also
necessary for getting sufficient precission in the structure
and electronic transition data. At this level, all valence
electrons of the \ec\
were correlated, which amounts to
86 and 85 electrons (Yb: 22 or 21; eight X: 64).
We used 
the standard IPEA parameter value (0.25 a.u.)
introduced
in Ref.~\onlinecite{GHIGO:04} as a simple way
to correct for systematic underestimations of
CASPT2 transition energies from closed-shell ground
states to open-shell excited states, also recommended
for other cases. An imaginary shift of 0.1 was also
used to avoid intruder states.~\cite{FORSBERG:97}


The highly correlated wavefunctions and
energies resulting from the previous spin-orbit
free step, namely, 
the eigenvectors (which are transformed RASSCF wave functions)
and eigenvalues (MS-RASPT2 energies)
of the spin-orbit free effective Hamiltonian computed
at the MS-RASPT2 step, were transferred and used in the last step of
spin-orbit coupling calculations, which used the
restricted-active-space state interaction method (RASSI-SO).~\cite{MALMQVIST:02}
The transformed RASSCF wave functions
were used to compute the spin-orbit coupling matrix resulting from
adding the AMFI approximation of the Douglas-Kroll-Hess spin-orbit
coupling operator~\cite{HESS:96}
to the scalar relativistic Hamiltonian.
Given that \Dth\ symmetry
was used, the MS-RASPT2 energies
show slight degeneracy breakings which would spread further throughout
the spin-orbit calculation. Hence, the spin-free-state-shifting
technique was used to substitute them for averaged values
which were used in the diagonal elements of the spin-orbit coupling matrix.
The calculation and diagonalization of the 
transformed RASSCF spin-orbit interaction matrix 
leads to the final results of the \abinitio\ calculations on the
independent \Dclus\ and \Aclus\ active centers, which
can be used in Eq.~\ref{EQ:E02} to compute the
diabatic potential energy surfaces of the \Dclus--\Aclus\
mixed valence active pairs.

The program MOLCAS was used for all calculations.~\cite{MOLCAS}
All AIMP data (for embedding and/or for cores) and
interstitial basis set can be found in
Ref.~\onlinecite{AIMP-DATA}.

\subsubsection{Energy curves of \Ybtp\ and \Ybthp\ independent active centers}
\label{SEC:YbIIres}

The results of the quantum mechanical calculations on the 
donor
\Dclus\ and acceptor \Aclus\ cubic clusters embedded in the four MX$_2$ hosts
are collected in the Suplementary Material of this 
paper,
where the details of the electronic structure of the
ground and excited states of the \Ybtp\ and \Ybthp\ active centers
are presented.~\cite{SupMatRef}
Tables I to IV of Ref.~\onlinecite{SupMatRef}
include the Yb--X equilibrium bond distance, totally
symmetric vibrational frequency, minimum-to-minimum energy
differences relative to the $4f^{14}$--1\Aog\ ground state, 
and analyses of the spin-orbit wavefunctions
of all 7 \Aou, 7 \Atu, 14 \Eu, 21 \Tou, and 21 \Ttu\ spin-orbit
levels calculated. 
They also include the results for the electronic states of the $4f^{13}$ configuration
of the \Ybthp\ active centers.
Plots of their energy curves $vs.$ the
Yb--X bond distance at different levels of methodology
(RASSCF, MS-RASPT2, and RASSI-SO) and the calculated absorption spectra
are also presented in 
Figures 1 to 5 
of Ref.~\onlinecite{SupMatRef}.
The data of only the lowest levels (14  of \Ybtp\ and 5 
 of \Ybthp) in CaF$_2$ are presented here in
Table~\ref{TAB:Ca-SpinOrbit-short};
selected energy curves are presented in the left graphs of Fig.~\ref{FIG:monomers-dimer};
all the energy curves are presented in the left grpah of Fig.~\ref{FIG:CaF2-monomers-dimer}.

The excited states of \Ybtp\ in the four hosts appear to be
grouped in four manifolds of impurity states (see Table~\ref{TAB:Ca-SpinOrbit-short} 
and left graph of Fig.~\ref{FIG:CaF2-monomers-dimer} for CaF$_2$):
$4f^{13}(7/2)5de_g$,
$4f^{13}(5/2)5de_g$,
$4f^{13}(7/2)5dt_{2g}$, and
$4f^{13}(5/2)5dt_{2g}$, and
two of impurity-trapped excitons (ITE):
$4f^{13}(7/2)a_{1g}^{YbTE}$ and
$4f^{13}(5/2)a_{1g}^{YbTE}$, which, in some cases,
show significant configurational interaction.
The more delocalized 
ITE states, whose electronic structure shows the same characteristics as
those found in previous quantum mechanical studies on
\Ufp-doped Cs$_2$GeF$_6$,~\cite{ORDEJON:07}
and \Ybtp-doped SrCl$_2$,~\cite{SANCHEZ-SANZ:10:a}
appear above the lowest $4f^{13}(7/2)5de_g$ impurity manifold in
all four hosts. 
The lowest excited states of \Ybtp\ are, in all cases,
the $4f^{13}(7/2)5de_g$ electric dipole forbidden  1\Eu, 1\Ttu, 
and the electric dipole allowed   1\Tou\ states.
(In all figures, blue and green colours have been used for the energy curves of the
$4f^{13}5de_g$ and $4f^{13}5dt_{2g}$, manifolds, respectively; maroon has been used
when they interact among themselves or with ITE states. Black has been used for the
\Ybtp\ ground state and for the lowest $4f^{13}(7/2)$ manifold of \Ybthp; red is
used for the higher $4f^{13}(7/2)$ components.)

The present \abinitio\ results and those of calculations presented elsewhere,~\cite{BARANDIARAN:TBP}
especially
designed to allow the wavefunctions of excited states to spread electron 
density beyond first neighbours, over the twelve next \Catp\ or \Srtp\ ions,
as proposed by McClure and P\'edrini for impurity-trapped excitons,~\cite{MCCLURE:85,MOINE:89} 
allow to conclude that none of the electronic states of the \Ybtp\ active centers can be
considered responsible for
the anomalous emission observed in the CaF$_2$ and SrF$_2$ hosts.
Therefore,     the hypothesis of the impurity-trapped
excitons being responsible for the anomalous emission~\cite{MCCLURE:85,MOINE:89} is not supported by the
\abinitio\ quantum mechanical calculations.
  Impurity-trapped excitons show up in the calculations,
  but at much higher energies than those that would allow them to play a role in the anomalous emission.
As we show in Sec.\ref{SEC:IVCT-CaF2} and \ref{SEC:interplay}, intervalence charge transfer within the 
\Ybiipiiip\ active pairs has to be invoked instead.

\subsection{Quantitative energy diagrams for \Ybiipiiip\ intervalence charge transfer reaction}
\label{SEC:energy-diagrams}

Even though the calculation of the two diabatic potential energy surfaces
$E_{DiAj}^{\rm diab}(d_L,d_R)$ and $E_{AjDi}^{\rm diab}(d_L,d_R)$ associated
with each $ij$ state
of the embedded pair before and after electron transfer is necessary
for the obtention of the structural and energetic results discussed in this
paper (note that any of the 
MX$_2$:\Ybiipiiip\ excited states lead to two energy surfaces like those
for the ground state presented in 
Fig.~\ref{FIG:ivct-gs-Qet}, 
which would 
make their joint plot extremely crowded and cumbersome),
the extraction of the minimal energy electron transfer reaction
path from them is a very important step leading to 
quantitative IVCT energy diagrams which will be actually used
to discuss IVCT luminescence and other radiative and non-radiative IVCT
processes in Sec.~\ref{SEC:IVCT-CaF2} and ~\ref{SEC:interplay}. In this Section we explain how these diagrams are built:
In a first step, the normal electron transfer reaction coordinate \Qet\ for the
ground state of the mixed valence pair, \Qetzz, will be defined from the topology
of the $E_{D0A0}^{\rm diab}(d_L,d_R)$ and $E_{A0D0}^{\rm diab}(d_L,d_R)$ energy surfaces, in terms of 
concerted vibrational
breathing modes of the donor $D$ and acceptor $A$ sites, which leads to 
the diabatic $E_{D0A0}^{\rm diab}$(\Qetzz) 
and $E_{A0D0}^{\rm diab}$(\Qetzz) branches for the reaction.
In a second step, the energy curves of the excited states of 
the mixed valence pairs will be evaluated at the 
$d_L$, $d_R$ coordinate values which correspond to the \Qetzz\
axis, leading to $E_{DiAj}^{\rm diab}$(\Qetzz) and $E_{AjDi}^{\rm diab}$(\Qetzz)
branches for all states of interest and, hence, to the final quantitative IVCT diagram.
The same procedure would be applicable to any other two
$E_{DiAj}^{\rm diab}(d_L,d_R)$ and $E_{AjDi}^{\rm diab}(d_L,d_R)$
diabatic surfaces, which would lead to analogous
energy diagrams along corresponding \Qetij\ reaction coordinates.

\subsubsection{Electron transfer reaction coordinate}
\label{SEC:Qet}

Within the diabatic approximation,
the two symmetric diabatic energy surfaces involved in the
electron transfer reaction between the $D$ and $A$ ground states,
$E_{D0A0}^{\rm diab}(d_L,d_R)$ and  $E_{A0D0}^{\rm diab}(d_L,d_R)$,
do intersect, without avoiding crossing, at
all $d_L = d_R$ points (see 
Fig.~\ref{FIG:ivct-gs-Qet}), 
where they are degenerate, while the corresponding pair
wavefunctions $M\hat A (\Phi_{D0} \Phi_{A0})$ and $M\hat A (\Phi_{A0} \Phi_{D0})$
never mix, but keep their pure character. 
In this way, the
electron transfer reaction path  
(see Fig.~\ref{FIG:ivct-gs-Qet}, top right graph)
connects the energy minimum of 
the $DA$ \DclusL--\AclusR\ pair, $E_{e,00}$=$E_{D0A0}^{\rm diab}(d_{eD},d_{eA})$,
for which the \DclusL\ and \AclusR\ clusters are at their Yb--X equilibrium distance,
$d_L=d_{eD}$ and $d_R=d_{eA}$, with that of the symmetrical minimum of the
$AD$ \AclusL--\DclusR\ pair, $E_{e,00}$=$E_{A0D0}^{\rm diab}(d_{eA},d_{eD})$,
at the $d_L=d_{eA}$, $d_R=d_{eD}$ point,  crossing from the $DA$ to 
the $AD$ surface through the activated complex, 
which is the crossing point
with minimal energy, equal sharing of the transferred
electron, and equal \mbox{Yb--X} bond distance for the left and right clusters of the
$DA$ and $AD$ pairs, $d_L=d_R=d_{ac}$.
The diabatic electron transfer activation energy:
$E_{D0A0}^{\# \rm diab} = E_{D0A0}^{\rm diab}(d_{ac},d_{ac}) - E_{e,00}$,
is an upperbound to the energy barrier that would result from avoided crossings
should the electronic coupling between the  two energy surfaces be considered.
Whereas the diabatic activation energy is independent from the distance between
the $D$ and $A$ moieties of the pairs, $d_{DA}$, the adiabatic energy barrier is
$d_{DA}$-dependent.
As commented above, 
it is close to the activated complex structure where the diabatic 
approximation is weakest.

The ground state diabatic reaction coordinate $Q_{et}$
(for simplicity we will drop the superscript in \Qetzz\ from now on) 
can be aproximated with the straight lines that connect the
activated complex $(d_{ac},d_{ac})$ with the two minima $(d_{eD},d_{eA})$ and $(d_{eA},d_{eD})$.
This reaction coordinate is represented in 
Fig.~\ref{FIG:ivct-gs-Qet} 
in the $d_L$-$d_R$ plane.
The $DA$ pair has the lowest diabatic energy in the left side of the activation complex ($d_L > d_R$)
and the $AD$ pair in the right side ($d_L < d_R$),
in correspondance with the larger size of $D$ at equilibrium.
Since these lines contain the most interesting information of the diabatic energy surfaces,
it is convenient to plot them in energy diagrams along the reaction coordinate:
$E_{DiAj}^{\rm diab}$(\Qet) and $E_{AjDi}^{\rm diab}$(\Qet)
instead of the more cumbersome $(d_L,d_R)$--dependent energy surfaces,
$E_{D0A0}^{\rm diab}(d_L,d_R)$ and  $E_{A0D0}^{\rm diab}(d_L,d_R)$, as mentioned above.

For a precise definition of $Q_{et}$, we can recall that
the changes of the Yb--X distances in the left and right clusters $d_L$ and $d_R$ 
along the reaction coordinate fulfil
\begin{eqnarray}
  d_L-d_{ac} = m(d_R-d_{ac}) \;\left\{ 
    \begin{array}{cl} 
        m=\frac{d_{eD}-d_{ac}}{d_{eA}-d_{ac}} & : d_L \ge d_R \\ 
        m=\frac{d_{eA}-d_{ac}}{d_{eD}-d_{ac}} & : d_L \le d_R
    \end{array} \right.
\,.
\label{EQ:E04}
\end{eqnarray} 
Then, the normal reaction coordinate can be written as
\begin{eqnarray}
  Q_{et} = \frac{1}{\sqrt{1+m^2}} \left( Q_R + m Q_L \right)
\,,
\end{eqnarray} 
$Q_L$ and $Q_R$ being the normal breathing modes of the left and right YbX$_8$ moieties
with respect to their structures in the activated complex:
\begin{eqnarray}
  Q_L = \frac{1}{\sqrt{8}} \left( \delta_{X_{L1}} + \delta_{X_{L2}} + \ldots +  \delta_{X_{L8}} \right) 
\,,
\nonumber\\
  Q_R = \frac{1}{\sqrt{8}} \left( \delta_{X_{R1}} + \delta_{X_{R2}} + \ldots +  \delta_{X_{R8}} \right) 
\,,
\end{eqnarray}
which have been expressed in terms of   
the displacements $\delta_{X_{Lk}}$ and $\delta_{X_{Rk}}$
of the X ligand atoms in the left and right YbX$_8$ moieties
away from their respective Yb atoms,
starting from the positions they occupy in the activated complex.
A graphical representation of the X displacements along $Q_{et}$ is shown in 
Fig.~\ref{FIG:ivct-gs-Qet} (bottom graph).
\iftoggle{journal-like}{
  \def\escalafig{0.55} 
  \begin{figure}[h!] 
    \input{FIG-\FIGIII-fig.tex}
    \input{FIG-\FIGIII-cap.tex}
  \end{figure}
}{}
Since the left and right cluster breathings imply
$\delta_{X_{L1}} = \delta_{X_{L2}} = \ldots = d_L - d_{ac}$ and
$\delta_{X_{R1}} = \delta_{X_{R2}} = \ldots = d_R - d_{ac}$,
we can write 
\begin{eqnarray}
  Q_L = \sqrt{8} (d_L - d_{ac})
\,,
\nonumber\\
  Q_R = \sqrt{8} (d_R - d_{ac})
\,,
\end{eqnarray} 
and
\begin{eqnarray}
  Q_{et} &=& \sqrt{\frac{8}{1+m^2}} \left[ (d_R - d_{ac}) + m (d_L - d_{ac}) \right]
\nonumber\\
         &=& \sqrt{8(1+m^2)} (d_R - d_{ac})
\nonumber\\
         &=& \sqrt{8(1+m^2)} (d_L - d_{ac})/m
\,.
\label{EQ:E08}
\end{eqnarray} 

We may mention the relationship of this reaction coordinate and the one of 
the vibronic model of Piepho \etal.~\cite{PIEPHO:78}
The latter correspons to $d_{ac} = (d_{eD} + d_{eA})/2$, which implies $m=-1$ and gives
 $Q_{et} = \frac{1}{\sqrt{2}}(Q_R-Q_L) = 2(d_R-d_L)$.

\subsubsection{Quantitative \Ybiipiiip\ IVCT energy diagrams}
\label{SEC:build-IVCT-diagrams}

According to Eq.~\ref{EQ:E02} and Eqs.~\ref{EQ:E04}--\ref{EQ:E08},
quantitative IVCT energy diagrams for the embedded
\Ybiipiiip\ active pairs can be obtained  using the energy curves
resulting from the independent embedded cluster calculations
of the electronic states of the donor \Dclus\ and acceptor \Aclus\
active centers described in Sec.~\ref{SEC:iecmeth}.
We illustrate here how this is done using 
the energy curves of only two electronic states for the
donor and acceptor clusters embedded in CaF$_2$,
($n_D=2, n_A= 2$):
the ground state and one excited state,
which have been plotted in
the left graph of 
Fig.~\ref{FIG:monomers-dimer}. The IVCT energy diagram resulting from them, which includes
$2n_Dn_A= 8$ diabatic mixed valence pair states, is
plotted in the right graph of Fig.~\ref{FIG:monomers-dimer} and is
described next. 
The much more dense IVCT energy diagrams used below in this paper result from
obvious extensions which use the complete manifold of excited states calculated
for the separated active centers.

Using the energy curves of the 1\Aog\ and 1$\Gamma_{7u}$ ground states of the
\Ybtp\ and 
\Ybthp\ doped CaF$_2$ (see Table~\ref{TAB:Ca-SpinOrbit-short}),
the diabatic potential energy surfaces $E_{D0A0}^{\rm diab}(d_L,d_R)$ and  $E_{A0D0}^{\rm diab}(d_L,d_R)$,
for the pair before \DclusL--\AclusR\ [1\Aog,1$\Gamma_{7u}$] and  
after \AclusL--\DclusR\ [1$\Gamma_{7u}$,1\Aog] electron transfer,
are obtained from Eq.~\ref{EQ:E02}.
From any of them, the activated complex structure, $d_{ac}$=2.261~\AA, 
is obtained searching for the
minimal energy along the $d_L=d_R$ axis. 
Using this value, plus the Yb--F equilibrium distances of the
independent embedded cluster ground states, \mbox{$d_{eD}$=2.329~\AA\ (1\Aog)} and 
\mbox{$d_{eA}$=2.201~\AA} (1$\Gamma_{7u}$), respectively, the diabatic 
electron transfer reaction coordinate
for the ground state of the mixed valence pair is defined from Eqs.~\ref{EQ:E04}--\ref{EQ:E08},
and the two mixed valence pair ground state branches
[1\Aog,1$\Gamma_{7u}$]  and [1$\Gamma_{7u}$,1\Aog] 
 are obtained; they are plotted in black
in the right graph of Fig.~\ref{FIG:monomers-dimer}.
The two diabatic energy curves cross at the activated complex
structure (\Qet=0 \AA; $d_L=d_R=2.261$ \AA; $E_{D0A0}^{\# \rm diab}$=3774~\cmm1).
  In the structures of the pair around the minimum of the $DA$ state [1\Aog,1$\Gamma_{7u}$]
  (\mbox{\Qet= --0.256~\AA;}$d_L=d_{eD}$=2.329, $d_R=d_{eA}$=2.201 \AA),
  the complementary $AD$ state [1$\Gamma_{7u}$,1\Aog] has a much higher energy 
  because the oxidized and reduced moieties, $A$ and $D$, are under strong structural stress. 
  The opposite is true around the minimum of the $AD$ state [1$\Gamma_{7u}$,1\Aog].
  We indicate this in the energy diagram by representing the energy of a given mixed valence pair state
  with full lines when it is around its minimum,
  and with dashed lines when it is under strong stress.
  Computing the energies of the excited states of the pairs at the $d_L,d_R$ values corresponding to
  the ground state electron transfer coordinate we obtained the energy curves of the $DA$ states
  [1\Aog,2$\Gamma_{7u}$] (red), [1\Eu,1$\Gamma_{7u}$] (blue), and [1\Eu,2$\Gamma_{7u}$] (orange),
  plus their IVCT $AD$ counterparts.
  They are plotted in the right graph of Fig.~\ref{FIG:monomers-dimer} 
  using the above definition of solid and dashed lines.

The quantitative IVCT diagram shown in Fig.~\ref{FIG:monomers-dimer} 
can be used to read different types of processes.

The $D_0A_0 \rightarrow A_0D_0$ thermally induced ground state intervalence electron transfer reaction, 
\Ybiipiiip\ [1\Aog,1$\Gamma_{7u}$] $\rightarrow$ \Ybiiipiip\ [1$\Gamma_{7u}$,1\Aog],
with an upperbound 
activation energy barrier of 3774~\cmm1\ (black, solid line
connecting the two equivalent ground state
minima through the activated complex).

Vertical absorptions of the independent active centers, 
which occur at their (fixed) ground state structure, such as:
the \Ybthp\ $4f$-$4f$ absorption $1\Gamma_{7u}\rightarrow 2\Gamma_{7u}$, 
at a Yb$^{3+}$-F distance 2.201~\AA\  ($D_0A_0 \rightarrow D_0A_1$, arrow 1), 
and the \Ybtp\ $4f$-$5d$ absorption $1A_{1g}\rightarrow 1E_u$, 
at a Yb$^{2+}$-F distance 2.329~\AA\ ($D_0A_0 \rightarrow D_1A_0$, arrow 2).
These transitions are the same read in the energy curves of 
the independent active centers in the left graph of Fig.~\ref{FIG:monomers-dimer}.

Vertical IVCT absorptions from the pair ground state, 
at fixed \mbox{\Qet=--0.256~\AA}, which means a left Yb-F distance 2.329~\AA\ and a right Yb-F distance 2.201~\AA.
The initial pair \DclusL--\AclusR\ is relaxed and, after the radiatively induced electron transfer,
the final pair \AclusL--\DclusR\ is very stressed and far from equilibrium.
These absortion bands are very wide.
There is a lowest IVCT absorption ($D_0A_0 \rightarrow A_0D_0$, arrow 3), 
which leads to the photoinduced ground state electron transfer reaction after nonradiative decay,
and higher lying IVCT absorptions, like $D_0A_0 \rightarrow A_0D_1$ (arrow 4).

Electron transfer non-radiative decays to the ground state or to excited states of the pair
can also be visualized in the IVCT energy diagram,
like the $D_0A_1 \rightarrow A_0D_0$ or the $D_1A_0 \rightarrow A_1D_0$ decays.
In this cases, although the IVCT diagram is very helpul identifying the most probable decay mechanisms,
the corresponding energy barriers have to be found in the full diabatic potential energy surfaces involved,
$E_{DiAj}^{\rm diab}(d_L,d_R)$ and $E_{AkDl}^{\rm diab}(d_L,d_R)$, 
because they may occur at \dL,\dR\ points
not corresponding to the ground state electron transfer reaction coordinate \Qet\ axis. 

Finally,
vertical emissions can also be discussed using the IVCT energy diagrams
in analogous terms as we have discussed vertical absorptions above.
These can be regular emissions of the independent acive centers, like $A_1D_0 \rightarrow A_0D_0$ (arrrow 5),
and IVCT emissions, like $A_1D_1 \rightarrow D_0A_0$ (arrow 6).
Rigorously, the electron transfer reaction coordinate \Qet\ axis should
correspond to the $ij$ emitting state of the pair. Yet, the IVCT diagram
obtained for the \Qetzz\ and \Qetij\ axes are very similar in most
cases. Hence, although the transition energies tabulated in this work 
will be calculated rigorously, we will use the ground state IVCT energy diagram to 
discuss absorption, decays, and emission processes altogether.
Therefore, independent active center emissions and IVCT emissions
will be discussed in analogous terms as the independent active center
absorptions and IVCT absorptions, with obvious substitutions of the \Qet\ fixed values
involved in the radiative processes and their effects on the
emission band structure.

\section{IVCT luminescence in CaF$_2$:\Ybtp}
\label{SEC:IVCT-CaF2}

  In this Section we present and discuss in detail the IVCT luminescence of CaF$_2$:\Ybtp.
   The manifolds of excited states of \Ybiipiiip\ pairs in CaF$_2$ are introduced in Sec.~\ref{SEC:es-CaF2}
   and they are used to interpret the anomalous luminescence of CaF$_2$:\Ybtp\ as an IVCT luminescence in Sec.~\ref{SEC:mechanism},
   where its mechanism is discussed in detail. 
   Additionally, we discuss two experiments that reinforce the involvement of IVCT states:
   Firstly, in Sec.~\ref{SEC:IRESA}  
   we show that the transient photoluminescence enhancement measured via two-frequency excitation experiments
   can be interpreted, in detail, as an IVCT photoluminescence enhancement experiment; 
   then, in Sec.~\ref{SEC:YbIIILumin} we show that the IVCT model also explains 
   the excitation of the IR \Ybthp\ emission by \Ybtp\ \fdabs\ absorptions.

  \subsection{Excited states of \Ybiipiiip\ pairs}
  \label{SEC:es-CaF2}

The full diabatic IVCT energy diagram for the ground and excited states of the
\Ybiipiiip\ pairs in CaF$_2$, along the normal
reaction coordinate of the ground state, can be seen in
Fig.~\ref{FIG:CaF2-monomers-dimer}. It has been built using all the parent 
independent embedded cluster energy curves plotted in the left graph, following the procedure
explained in Section 
\ref{SEC:build-IVCT-diagrams}.
The vertical transition energies from the \Ybiipiiip\ [1\Aog,1\Gsevu]  ground state 
minimum (\mbox{$Q_{et}$= --0.256,} \dL,\dR= 2.330,2.201~\AA; zero energy in the diagram),
which can be read directly from the diagram, are also collected in Table~\ref{TAB:CaSrF2-YbIIYbIII}.
As indicated in the Table, only the excited states which are relevant for the discussions
of this paper have been tabulated; the remaining data is 
available in Ref.~\onlinecite{SupMatRef}.
Vertical transition energies from the minima of other two electronic states of the
\Ybiipiiip\ pairs, [1\Aog,2\Gsevu] and [2\Aou,1\Gsevu], have also been  included in Table~\ref{TAB:CaSrF2-YbIIYbIII}.
In the  [1\Aog,2\Gsevu] state the \Aclus\ right moiety of the \Ybiipiiip\ pair is excited
in the lowest of the $4f^{13}$(5/2) levels;
in the [2\Aou,1\Gsevu] state the \Dclus\ left moiety is excited in the
lowest of $4f^{13}(5/2)5de_g$ levels (cf. symmetry labels and configurational character of the parent independent embedded cluster
states in Table~\ref{TAB:Ca-SpinOrbit-short}). Vertical absorptions and emissions from their minima will be
used in the next subsections; however, their
values cannot be read directly from the IVCT diagram of Fig.~\ref{FIG:CaF2-monomers-dimer},
since its electron transfer reaction coordinate is that of the ground state. 
Specific IVCT diagrams along the [1\Aog,2\Gsevu] and [2\Aou,1\Gsevu]
reaction coordinates would have to be constructed using Eqs.~\ref{EQ:E04}--\ref{EQ:E08}
and the procedure outlined in Sec.~\ref{SEC:build-IVCT-diagrams}. However, the similarity of all
three IVCT diagrams allows to visualize the data from Table~\ref{TAB:CaSrF2-YbIIYbIII}
in Fig.~\ref{FIG:CaF2-monomers-dimer}.

Diabatic energy barriers for the IVCT reaction \Ybiipiiip\ $\rightarrow$ \Ybiiipiip,  connecting two different 
electronic states of the pairs: \mbox{[$D_i,A_j$]  $\rightarrow$ [$A_k,D_l$]}, have been
included in Table~\ref{TAB:IVCTbarr} for the forward and backwards reactions,
together with the \dL,\dR\ coordinates of their
activated complex point,  since they play an
important role in non-radiative decay pathways, as discussed below.
The forward/backwards energy barriers have been calculated searching for the activated complex point of the
[$D_i,A_j$]  $\rightarrow$ [$A_k,D_l$] reaction directly along the intersection points of the two
$E_{DiAj}^{\rm diab}(d_L,d_R)$ and $E_{AkDl}^{\rm diab}(d_L,d_R)$
energy surfaces. Again, even though the activated complex \dL,\dR\ structures 
gathered in the Table do not necessarily
fall in the ground state
electron transfer reaction coordinate of Fig.~\ref{FIG:CaF2-monomers-dimer}, 
they are not far from the crossing points of their respective 
$E_{DiAj}^{\rm diab}$(\Qet)
and $E_{AkDl}^{\rm diab}$(\Qet) branches;
so, they can be visualized in Fig.~\ref{FIG:CaF2-monomers-dimer}.

In the following and for clarity, the IVCT energy diagram of Fig.~\ref{FIG:CaF2-monomers-dimer} will be reduced
so as to include only the electronic states that have been found to participate in the
IVCT luminescence mechanism of Yb-doped CaF$_2$ crystals 
and in the electronic spectroscopy experiments that will be
discussed below.
The reduced IVCT diagram  appears in Fig.~\ref{FIG:CaF2-YbIIYbIII-11-few}.
 
  \subsection{IVCT luminescence mechanism}
  \label{SEC:mechanism}

The results of the quantum mechanical calculations described in the preceding sections
suggest that the IVCT luminescence, which has been observed experimentally and has been
interpreted so far as an $anomalous$ luminescence, occurs according to the
following mechanism involving steps I to V. 
The data which characterize the radiative and
non-radiative processes the \Ybiipiiip\ pair states undergo, can be found in Table~\ref{TAB:CaSrF2-YbIIYbIII} 
and Fig.~\ref{FIG:CaF2-YbIIYbIII-11-few}; Table~\ref{TAB:Ca-SpinOrbit-short} should
be useful to clarify the electronic structure of their \Ybtp\ and \Ybthp\ parent states.

  \subsubsection{Step I. First photon absorption.}
  Step I is the first photon absorption {[1\Aog,1\Gsevu] $\rightarrow$ [1\Tou,1\Gsevu]}.

  This step is  the lowest electric dipole allowed $4f^{14}$ $\rightarrow$ $4f^{13}5de_g$ 
  excitation of \Ybtp: \mbox{1\Aog\ $\rightarrow$ 1\Tou}.~\cite{LOH:68,LOH:69}
  Temperature dependent (multiphonon) non-radiative decay to the lowest lying state of the $4f^{13}5de_g$ configuration, 1\Eu,
  can be expected to occur.~\cite{GRIMM:05,PAN:08}
  Alternatively, a lower energy photon can excite directly the 1\Eu\ state with a less efficient, electric dipole forbidden transition.~\cite{MOINE:89}
  These two states can be labelled as the \mbox{[1\Tou,1\Gsevu]} and \mbox{[1\Eu,1\Gsevu]} states of the \Ybiiipiip\ pair, respectively.
  Comparisons between their  calculated vertical transition energies and experimental peak energies~\cite{FEOFILOV:56,KAPLYANSKII:62,LOH:69}
  (Table~\ref{TAB:CaSrF2-YbIIYbIII}, 25706~\cmm1\ and 23576~\cmm1\ vs. 27400~\cmm1\ and 24814~\cmm1, respectively),
  suggest similar overestiamtions of around 1500~\cmm1.

  \subsubsection{Step II. Non-radiative electron transfer} 

  Step II is the non-radiative electron transfer
  [1\Tou,1\Gsevu] and/or [1\Eu,1\Gsevu] $\rightarrow$ [2\Gsevu,1\Aog].

  A very small energy barrier (49~\cmm1) is found for  
  \Ybiipiiip~[1\Tou,1\Gsevu] to \Ybiiipiip~[2\Gsevu,1\Aog]
  electron transfer,
  which suggests that such non-radiative decay is very likely to ocurr. 
  At the activated complex point of the electron transfer, (\dL,\dR)=(2.323~\AA, 2.197~\AA)
  the resulting \Ybiiipiip\ [2\Gsevu,1\Aog] state is so structurally stressed
  that a profound non-radiative relaxation towards its final equilibrium structure
  (\dL,\dR)=(2.201~\AA, 2.330~\AA) follows (see step II in Fig.~\ref{FIG:CaF2-YbIIYbIII-11-few}).
  The same is true for the [1\Eu,1\Gsevu] $\rightarrow$ [2\Gsevu,1\Aog]
  non-radiative charge transfer for wich the energy barrier is found to be
  even smaller: 14~\cmm1.

  These small energy barriers suggest that step II,
  which is an
  \YbII($^2F_{7/2}5de_g$)--\YbIII($^2F_{7/2}$) $\rightarrow$ \YbIII($^2F_{5/2}$)--\YbII($4f^{14}$)
  non-radiative IVCT that leaves \YbIII\ in its $^2F_{5/2}$~~2\Gsevu\ excited state,
  should be an efficient quenching mechanism for regular $4f^{13}5de_g$$\rightarrow$$4f^{14}$ radiative emissions back to the
  ground state. This  explains that experimental detection of these emissions has never been reported
  for the CaF$_2$ host, as far as we know.

  Going into more details, it is worth noticing that 
  [1\Eu,1\Gsevu] can excite the IVCT directly through non-radiative decay along step II, followed by steps III and IV explained below. 
  [1\Tou,1\Gsevu], however, can excite it directly through its own step II, but also indirectly through an intermediate decay to  [1\Eu,1\Gsevu].
This explains the difference observed experimentally in the short time part of the
intensity decay curves of the anomalous (IVCT) emission: whereas the intensity of the emission shows a risetime when 1\Tou\ is excited,
the risetime dissappears when 1\Eu\ is directly excited instead.~\cite{MOINE:89}

  The calculatons show branching of the step II non-radiative decays,
  which could result in quenching of the IVCT luminescence.
  Besides, the energy barrier for the electron transfer reaction back to the ground state of the \Ybiipiiip pair after step II,
  \mbox{\Ybiiipiip~[2\Gsevu,1\Aog]} $\rightarrow$ \mbox{\Ybiipiiip~[1\Aog,1\Gsevu]},
  is found to be 285~\cmm1.
  All this indicate that:
  (i) both non-radiative decays below the branching should occur, and
  (ii) the pairs that decay to the \mbox{\Ybiiipiip~[2\Gsevu,1\Aog]} minimum (the end of step II) 
  and can yield IVCT luminescence after the next steps,
  will still face temperature dependent decay to the ground state through the low energy barrier.
  This is in agreement with the fact that quenching of the anomalous luminescence of CaF$_2$:\Ybtp\
  has been observed at 180~K.~\cite{MOINE:89,RUBIO:91}
  Also, since the temperature dependence of the non-radiative decay from 
  the 1\Tou\ to the 1\Eu\ states of \YbII\ mentioned in step I,
  [1\Tou,1\Gsevu] $\rightarrow$  [1\Eu,1\Gsevu], 
  should influence step II as well, right at the initiation of the IVCT emission mechanism,
  the overall temperature dependence of the IVCT luminescence should be complex;
  this is further discussed in Sec.~\ref{SEC:CaF2SrF2}.

  \subsubsection{Step III. Second photon absorption.}
  Step III is the second photon absorption [2\Gsevu,1\Aog] $\rightarrow$ [2\Gsevu,1\Tou].

This step is equivalent to step I, since
\Ybtp\ is excited to its lowest
$4f^{14}$ $\rightarrow$ $4f^{13}5de_g$ electric dipole allowed level 1\Tou\ by
a second photon of the same wavelength as that of step I.
The difference is that now, after step II, the \Ybthp\ component of the pair
is in its 
 $^2F_{5/2}$~~2\Gsevu\ 
excited state instead of its
 $^2F_{1/2}$~~1\Gsevu\ 
ground state. 
As commented in step I, alternatively, the lower lying
[2\Gsevu,1\Eu] state can be directly excited in this step
by a second, lower energy photon.

This step reveals that the excitation of the IVCT luminescence is a 
two-photon process. Whether excitation of the yellow-green anomalous luminescence
is a one- or a two-photon process has not been investigated
experimentally, as far as we know.

  \subsubsection{Step IV. Non-radiative decay to the luminescent level.}
  Step IV is the non-radiative decay to the luminescent level [2\Gsevu,1\Tou] $\rightarrow$ [1\Gsevu,2\Aou].

The non-radiative relaxation from [2\Gsevu,1\Tou] to the lowest
[1\Gsevu,2\Aou] level in this energy region separated from lower lying states
by a large energy gap, is now significantly
different from that described in step I after excitation to the
[1\Tou,1\Gsevu] state with the first photon, because
the number and nature of the electronic states that are found below 
the excited level are now  different. 
In effect,  the first photon excites to the lowest levels of the \Ybiipiiip\ 
[$^2F_{7/2}5de_g$,$^2F_{7/2}$] 
manifold,
in which both the $4f^{13}$ inner-shell of \YbII\ and the $4f^{13}$ shell of \YbIII\ are in their $^2F_{7/2}$ ground spin-orbit multiplet.
However,
the second photon excites to a more crowded energy region, 
where two manifolds of the \Ybiiipiip\ pair share the same energy range:
  [$^2F_{7/2}$,$^2F_{5/2}5de_g$] and [$^2F_{5/2}$,$^2F_{7/2}5de_g$];
these two manifolds have their
$4f^{13}$ inner-shell of \YbII\ and $4f^{13}$ shell of \YbIII, respectively,
excited 
into the $^2F_{5/2}$ spin-orbit multiplet.
As a result, all of these energy levels lie close in energy (cf. Table~\ref{TAB:CaSrF2-YbIIYbIII})
and
non-radiative relaxation should not be hindered by energy gaps and 
it should be fast until the luminescent level [1\Gsevu,2\Aou] is reached.
This step would be reduced to a minimum when 
the [2\Gsevu,1\Eu] level is directly excited. 

  \subsubsection{Step V. IVCT luminescence.}
  Step V is the IVCT luminescence 
  [1\Gsevu,2\Aou] $\rightarrow$ [1\Aog,$^2F_{7/2}$].

In the last step, the IVCT luminescence consists of three vertical
\mbox{\Ybiiipiip\ $\rightarrow$ \Ybiipiiip} intervalence charge transfer transitions. 
They occur from the excited
\YbIII($^2F_{7/2}$)--\YbII($^2F_{5/2}5de_g$) state [1\Gsevu,2\Aou] 
 to the three components of the 
\YbII($4f^{14}$)--\YbIII($^2F_{7/2}$)  ground state:
[1\Aog,1\Geigu]  at 19974~\cmm1, 
[1\Aog,1\Gsixu]  at 19999~\cmm1, and
[1\Aog,1\Gsevu]  at 20508~\cmm1.
These leads basically to two wide bands 530~\cmm1\ apart.

The vertical transitions occur at the equilibrium structure of the initial [1\Gsevu,2\Aou] state:
(\dL,\dR)$_{e,{\rm initial}}$=(2.201~\AA, 2.314~\AA). 
This means that the \Ybiipiiip\ pair that results from the vertical electron transfer
is very much stressed, far away
from its own equilibrium structure at (\dL,\dR)$_{e,{\rm final}}$=(2.330~\AA, 2.201~\AA), which
defines the offsets on the left and right moieties of the pair as
$\Delta d_L$=+0.129~\AA\ and
$\Delta d_R$=--0.113~\AA,
and the offset in the normalized concerted vibrational electron transfer reaction coordinate
(the distance between  (\dL,\dR)$_{e,{\rm initial}}$ and  (\dL,\dR)$_{e,{\rm final}}$ in the \dL,\dR\ plane) 
as
$|\Delta Q_{et}|$ = $\sqrt{8}$ $\sqrt{(\Delta d_L)^2 + (\Delta d_R)^2}$ = 0.49~\AA.
This large offset results in very broad emission bands
with full width at
half maximum (FWHM) about 5800~\cmm1, much broader than it
could be expected from the regular \dFsh$5de_g$ $\rightarrow$ $4f^{14}$ emission, should
the latter occur.
This is illustrated  in Fig.~\ref{FIG:CaSrF2-IVCTL}, where the 
simulation of the IVCTL band has been produced for the three electronic origins
[1\Gsevu,2\Aou] $\rightarrow$ [1\Aog,1$\Gamma_{8u,6u,7u}$] using the semiclassical
time-dependent approach of Heller~\cite{HELLER:75,HELLER:81,ZINK:91:b} with an arbitrary value for the
oscillator strengths. 

The first reports on the luminescence of CaF$_2$:\Ybtp\ crystals below 200~K describe
a broad structureless band in the yellow-green region formed by two overlapping bands
peaking at 17600 and 18200~\cmm1,~\cite{FEOFILOV:56,REUT:76,MOINE:89} which suggests that the results of the 
calculations lead to overestimations of the IVCT luminescence by some 2000~\cmm1.
Values of the FWHM between 3000 and 4000~\cmm1\ have been found 
experimentally, depending on temperature.~\cite{MOINE:89}
The  2000~\cmm1\ overestimation of the calculated peaks of the IVCT bands, 
together with the 1700~\cmm1\ understimation obtained for 
the calculated \YbII\ $4f^{14}$--1\Aog\ $\rightarrow$ $4f^{13}5de_g$--1\Tou\ transition contribute to a
smaller calculated red shift of the
IVCT luminescence relative to the \Ybtp\ 
excitation than observed: it is found to be about 5600~\cmm1\ whereas the
experimentally measured red shift is about 10000~\cmm1.~\cite{RUBIO:91,MOINE:89}

An important characteristic of the level responsible for the
IVCT luminescence, [2\Aou,1\Gsevu] (or equivalently [1\Gsevu,2\Aou]),  
is its very low radiative rate compared with that of 
upper lying levels. This characteristic is basic for the efficiency of
transient photoluminescence enhancement experiments like the ones described
next.

  \subsection{Transient IVCT photoluminescence enhancement}
  \label{SEC:IRESA}

Two-frequency transient photoluminescence enhancement measurements on
CaF$_2$:\Ybtp\ single crystals at 10~K have been conducted to probe the energy levels
lying above the state responsible for  the so far called anomalous emission.~\cite{REID:11}
The experiments are based on the
"radically" different radiative decay rates of the lowest emitting state and 
higher excited states. The sample, excited in the UV at 365~nm (27400~\cmm1), is irradiated by
an IR pulse. The IR excitation induces significant enhancement of the emission because 
it populates
excited states 
that have significantly higher radiative rates.~\cite{REID:11} 
So, whereas the radiative rate of the lowest 
emitting level is in the order of 10$^1$ms, the decay of the transient
signal is much faster than 2$\times$10$^2$~$\mu$s.~\cite{REID:11}
  Here we show that these experiments can be interpreted as transient 
IVCT photoluminescence enhancement experiments.

Even though within the diabatic approximation used here we cannot calculate transition
moments between pair states and, hence, radiative rates or emission lifetimes, the analyses of 
the wavefunctions of the parent \Ybtp\ and \Ybthp\ states can give relevant
information on this subject.
In this line, the IVCT luminescence can be described as:
\mbox{
\Ybtp\ \dFfh$5de_g-2A_{1u}$
}
+ 
\mbox{
\Ybthp\ \dFsh$-1\Gamma_{7u}$
} 
$\rightarrow$
\mbox{
\Ybthp\ \dFsh$-1\Gamma_{7u}$
} 
 + 
\mbox{
\Ybtp\ $4f^{14}-1A_{1g}$
.}
This
indicates that the transition is a double 
orbital (or spinor) 
deexcitation, because \YbII\ gets its  $4f^{13}$ subshell deexcited 
(\YbII $4f_{7/2}$ $\rightarrow$ \YbII $4f_{5/2}$)
at the same time that its $5de_g$ electron is transferred to fill the \Ybthp\ hole
(\YbII $5de_g$ $\rightarrow$ \YbIII $4f_{7/2}$).
This is graphically represented in the scheme labeled double in Fig.~\ref{FIG:ivct-scheme}.
Analogous analysis of the emission from the next upper lying level [1\Eu,2\Gsevu]
yields the conclusion that its IVCT emission
\Ybiipiiip\ [1\Eu,2\Gsevu]
$\rightarrow$ 
\Ybiiipiip\ [2\Gsevu,1\Aog] 
is a single electron transfer
\mbox{\Ybtp\  $5de_g$ $\rightarrow$ \Ybthp\ $4f_{7/2}$} 
(see scheme labeled single in Fig.~\ref{FIG:ivct-scheme}).
Altogether, it is possible to conclude that the radiative rate of the 
lowest emission should be much lower than that of the next higher
emission, which is in agreement with the lowest emitting state having a long 
lifetime.~\cite{FEOFILOV:56,REUT:76,MOINE:89,REID:11}

Now, let us discuss the transient photoluminescence enhancement in more detail.
Analyses of the transient signals $vs.$ IR wavelength allowed to extract the
IR excited state absorption (ESA) spectrum originating in the lowest luminescent
level; it consists of two low-intensity sharp peaks at 250 and 1145~\cmm1\ and a 
higher intensity broad band from 650 to 950~\cmm1.~\cite{REID:11} Other signals found
were assigned to trap liberation processes and atmospheric absorptions
of the IR free electron laser FEL.~\cite{REID:11}
The calculated vertical IR ESA spectrum originating in the 
\Ybiipiiip\ [2\Aou,1\Gsevu] pair state has been included in Table~\ref{TAB:CaSrF2-YbIIYbIII}.
The simulation of the ESA spectrum of Fig.~\ref{FIG:CaSrF2-ESA}, graph (a), 
has been produced as the superposition
of narrow gaussians centered at the energy levels lying below  1700~\cmm1, using equal
values of the oscillator strength for each electronic transition. It also
consists of three groups of bands which correlate satisfactorily with the
experimental values from Ref.~\onlinecite{REID:11}, which have also been
included in Table~\ref{TAB:CaSrF2-YbIIYbIII} and 
Fig.~\ref{FIG:CaSrF2-ESA} for comparisons.
As in the previous paragraph, analyses
of the pair wavefunctions involved in the calculated IR ESA
spectrum in terms of the \Ybtp\ and \Ybthp\
parent states are useful to interpret the IR ESA signals observed, as follows:
The sharp experimental peak at 250~\cmm1\ corresponds to an energy 
transfer from the $4f^{13}$ subshell of \Ybtp\ to the $4f^{13}$ shell of  \Ybthp: 
$$
{\rm Yb}^{2+} 4f^{13}[4f_{7/2}\rightarrow 4f_{5/2}]5de_g
+ 
{\rm Yb}^{3+} 4f^{13}[4f_{5/2}\rightarrow 4f_{7/2}]
\,,
$$
or
\mbox{[$^2F_{5/2}5de_g$,$^2F_{7/2}$]} $\rightarrow$ \mbox{[$^2F_{7/2}5de_g$,$^2F_{5/2}$]} within the pair \YbII-\YbIII,
which is a \mbox{$4f$--$4f$} emission in \YbII\ and a \mbox{$4f$--$4f$} absorption in \YbIII.
In particular, the two IR ESA bands calculated at 270 and 307~\cmm1\ correspond to
 [2\Aou,1\Gsevu]$\rightarrow$[1\Eu,2\Gsevu] and  [2\Aou,1\Gsevu]$\rightarrow$[1\Ttu,2\Gsevu].
The sharp experimental peak at 1145~\cmm1\ corresponds to the $2A_{1u} \rightarrow 5T_{1u}$ intraconfigurational absorption
occurring in the \Ybtp\ center of the pair
\begin{center}
\Ybtp\ \dFfh$5de_g$$\rightarrow$\Ybtp\ \dFfh$5de_g$,
\end{center}
as it correlates well with the [2\Aou,1\Gsevu]$\rightarrow$[5\Tou,1\Gsevu] band calculated at 1069~\cmm1.
Finally, the broad band observed from 650 to 950~\cmm1\ is found to correspond to a set of
transitions calculated from 686 to 739~\cmm1, which span a narrower energy interval.
Here, higher  components of the energy transfer transitions from the 
$4f^{13}$ subshell of \Ybtp\ to the $4f^{13}$ shell of \Ybthp,  like those of the lowest IR ESA band,
are found together with a third type 
of excitations which are intraconfigurational $4f$--$4f$ transitions 
between the Stark components of the \Ybthp\ moiety: 
$$
{\rm Yb}^{3+} 4f_{7/2}\rightarrow {\rm Yb}^{3+} 4f_{7/2}
\,.$$ 
In all cases, the electronic transitions involve negligible structural reorganization,
which explains why they are very narrow bands.

  \subsection{Excitation  of the IR \Ybthp\ luminescence}
  \label{SEC:YbIIILumin}

A study of the optical properties of \Ybthp-doped CaF$_2$ crystals
after application of reducing methods such as \mbox{$\gamma$-irradiation} and 
annealing in hydrogen, was conducted in order to investigate the
\Ybthp/\Ybtp\ conversion in the CaF$_2$ host.~\cite{KACZMAREK:05}
The study showed the appearence
of the $4f^{14}$$\rightarrow$$4f^{13}5d$ absorption bands
characteristic of \Ybtp-doped CaF$_2$ in the UV absorption spectrum of the
treated samples.~\cite{KACZMAREK:05} 
Two weak overlapping emission bands were observed at room temperature at 565 and 540~nm
(17699 and 18519~\cmm1) in $\gamma$-irradiated CaF$_2$:30at.\% \Ybthp\ samples excited with 357~nm
(28011~\cmm1), which     were  identified as the yellow-green anomalous emission of
\Ybtp-doped CaF$_2$, which was observable even at room temperature in this case.
A reduction on the intensity of the EPR signals
associated with \Ybthp\ centers was observed after $\gamma$-irradiation
at temperatures ranging 10 to 100~K.
All of these features were taken as evidences of 
partial reduction of \Ybthp\  to \Ybtp\ in the CaF$_2$ host.~\cite{KACZMAREK:05}

Since the existence of \Ybiipiiip\ mixed valence pairs
in these samples appears to be a reasonable hypothesis,
the interpretation of the spectral features mentioned can be done 
in terms of the IVCT diagrams calculated in this
work as explained in the previous sections, where the first absorption
of \Ybtp\ and the IVCT luminescence mechanism have been described in detail. 
Yet, the type of samples used in Ref.~\onlinecite{KACZMAREK:05} brings
the opportunity to examine the quality of some of the approximations assumed in
the IVCT theoretical model used in
this paper. In effect,
it is worth noticing that departure from cubic site symmetry is far
more likely in as-grown CaF$_2$:\Ybthp\  than in CaF$_2$:\Ybtp\ crystals
due to necessary charge compensation in the former.~\cite{KIRTON:67}
[Associated with charge compensations, optical absorption signals between
10257 and 10995~\cmm1\ have been found in CaF$_2$:\Ybthp\ crystals 
(this work, cubic: 10763, 11196~\cmm1) and have
been associated with cubic, tetragonal, trigonal, and
rhombic sites;~\cite{KIRTON:67} they all correspond to  different splittings of the
 \dFfh\ excited multiplet.]
This is also true after the 
CaF$_2$:\Ybthp\ samples are subjected to reducing treatments.~\cite{KACZMAREK:05} 
Yet, the shape, number of bands, and peak positions of the IVCT luminescence
bands (so far anomalous bands) reported for CaF$_2$:\Ybthp\ reduced samples 
in Ref.~\onlinecite{KACZMAREK:05} and for 
CaF$_2$:\Ybtp\ crystals in Ref.~\onlinecite{FEOFILOV:56,KAPLYANSKII:62}, are basically identical:
two broad, structureless overlapping bands are observed peaking at
17699 and 18519~\cmm1, and at 17600 and 18200~\cmm1, respectively.
This fact suggests that the diabatic approximation proposed here and
the 
disregard of non-cubic splittings due to local charge compensation and/or
mutual interaction between the moieties of the \Ybiipiiip\ pairs
are reasonable theoretical bases and allow for the quantitative interpretation
of the main spectral features of the \Ybiipiiip\ mixed valence pairs
at the cost of independent embedded cluster calculations.

  In the reduced samples of Ref.~\onlinecite{KACZMAREK:05}, the
  $4f^{14}$$\rightarrow$$4f^{13}5d$ absorption bands of \Ybtp\
  were found in the excitation spectrum where the 980~nm (10204~\cmm1) IR emission of \Ybthp\ was monitored.
  The IVCT energy diagram (Fig.~\ref{FIG:CaF2-YbIIYbIII-11-few})
  provides a mechanism for the excitation of the \Ybthp\
  emission by the first absorption band of \Ybtp: 
  a generalization of step I, followed by steps II and III$_{em}$.
  E.g. the lowest \YbII\ $4f^{14} \rightarrow ^2F_{7/2}5de_g$ excitation $A_{1g} \rightarrow 1T_{1u}$ is setp I,
  and the next $A_{1g} \rightarrow 2-4T_{1u}$ excitations will be followed by nonradiative decays to $1T_{1u}$;
  the lowest \YbII\ $4f^{14} \rightarrow ^2F_{5/2}5de_g$ excitation $A_{1g} \rightarrow 5T_{1u}$ (second $4f \rightarrow 5d$ band) 
  will be followed by IVCT non-radiative decay 
  [$^2F_{5/2}5de_g$,1\Gsevu] $\rightarrow$ [1\Gsevu,$^2F_{7/2}5de_g$], followed by steps II and III$_{em}$;
  and similar arguments hold for the higher energy states.
Obviously, the \Ybthp\ emission, step III$_{em}$, competes with step III,
which excites the IVCT luminescence through a second photon absorption. 
However, its quenching ability is probably
small given than step III$_{em}$ is basically (within the independent
embedded cluster approximation) an \Ybthp\ $4f$-$4f$ electric dipole forbidden
transition whereas step III is an electric dipole \Ybtp\ $4f$-$5d$ allowed absorption.


 \section{Interplay between IVCT and $5d \rightarrow 4f$ emissions in CaF$_2$, SrF$_2$, BaF$_2$, and SrCl$_2$.}
 \label{SEC:interplay}

  As we have discussed, only the very broad yellow-green IVCT luminescence has been detected in CaF$_2$:\Ybtp.~\cite{FEOFILOV:56,REUT:76,MOINE:89}
  In SrF$_2$:\Ybtp, the equivalent ``anomalous'' emission has been detected in the red,~\cite{REUT:76,MCCLURE:85}
  but a second much narrower, blue emission band was also found experimentally and
  was interpreted as a regular \dfem\ emission from the metastable states 1\Eu\ and 1\Ttu\ of \Ybtp.~\cite{MOINE:88}
  And in BaF$_2$:\Ybtp\ there is no emission at all 
  up to 1.5$\mu$m after strong laser excitation in the \fdabs\ 
  band.~\cite{MOINE:89}
  The remaining luminescence combination, i.e. that only regular \dfem\ 
  emission is observed,
  is found in SrCl$_2$:\Ybtp.~\cite{WITZKE:73,PAN:08}
  The \dfem\ 
  emissions of SrF$_2$:\Ybtp\ and SrCl$_2$:\Ybtp\ are different: 
  an \Ybtp\ electric dipole allowed emission of higher energy, from 1\Tou, is present in SrCl$_2$:\Ybtp,
  which was not found in SrF$_2$:\Ybtp\ from 4.4~K to room temperature.~\cite{WITZKE:73,MOINE:88,PAN:08}
  In this Section we discussed the reasons for such a chemical dependence of the luminescence.
  Firstly, we discuss the CaF$_2$, SrF$_2$, BaF$_2$ chemical series in Sec.~\ref{SEC:CaSrBaF2}.
  Then, we discuss the SrF$_2$, SrCl$_2$ series in Sec.~\ref{SEC:SrF2SrCl2}.

  The results of the quantum mechanical calculations of the diabatic potential energy surfaces and 
  the full and selected IVCT energy diagrams for the ground and excited states of the \Ybiipiiip\ embedded pairs 
  in the CaF$_2$, SrF$_2$, BaF$_2$ hosts are presented 
  in 
Fig.~6 of Ref.~\onlinecite{SupMatRef}
and Fig.~\ref{FIG:CaSrBaF2-YbIIYbIII-11-few},
  an in the SrF$_2$, SrCl$_2$ hosts in 
Fig.~7 of Ref.~\onlinecite{SupMatRef}
and Fig.~\ref{FIG:SrF2Cl2-YbIIYbIII-11-few}.
  Vertical transition energies
  and energy barriers for IVCT reactions between pair states have been included 
  in Tables~\ref{TAB:CaSrF2-YbIIYbIII} and \ref{TAB:IVCTbarr}, respectively (descriptions of their content
  given in Sec.~\ref{SEC:es-CaF2} are valid and readily adaptable to SrF$_2$ and BaF$_2$ results). 
  More data can be found in Ref.~\onlinecite{SupMatRef}.

 \subsection{Luminescence  of \Ybtp-doped CaF$_2$, SrF$_2$, and BaF$_2$}
 \label{SEC:CaSrBaF2}

Probably associated with the increasing lattice volume 
going from CaF$_2$ (163.0~\AA$^3$, 0\%) to SrF$_2$ (194.7~\AA$^3$, 19\%) to BaF$_2$ (238.3~\AA$^3$, 46\%), 
the offset between the Yb--F equilibrium distance 
of the ground states of the donor \Ybtp\ and acceptor \Ybthp\
independent active centers (1\Aog\ and 1\Gsevu, respectively),
increases significantly
from 0.129~\AA\ (0\%) to 0.154~\AA\ (19\%), to 0.200~\AA\ (55\%), in absolute values.~\cite{SupMatRef}
Other properties change as well (breating mode vibrational frequencies, ligand field
effects, etc.), but their variations in the series are noticeably smaller.~\cite{SupMatRef}
The bond length offsets mentioned determine the values of the normal electron transfer reaction
coordinate (\Qet) at the minimum of the \Ybiipiiip\ pair ground state [1\Aog,1\Gsevu] and, therefore, 
the offsets between the two equivalent minima \Ybiipiiip\ [1\Aog,1\Gsevu] and
\Ybiiipiip\ [1\Gsevu,1\Aog] along the \Qet\ axis.
In the fluoride series this offset increases as:
$|\Delta Q_{et}|$ = 0.517~\AA\ (0\%, CaF$_2$), 0.616~\AA\  (+19\%, SrF$_2$), 0.802~\AA\ (+55\%, BaF$_2$), 
as it can be observed in 
Fig.~6 of Ref.~\onlinecite{SupMatRef}
and Fig.~\ref{FIG:CaSrBaF2-YbIIYbIII-11-few}. 
The different luminescent behaviour of Yb in the fluoride series stems from
this trend because, the increase in the offset between the two minima in the IVCT configuration coordinate diagram provokes a shift of 
the crossing points between the 
lowest \Ybiipiiip\ [\dFsh$5de_g$,\dFsh] states 
and the  stressed branches of the \Ybiiipiip\ IVCT states [$^2F_{7/2},4f^{14}$] and [$^2F_{5/2},4f^{14}$],
which act as two different non-radiative decay pathways after the first photon excitation:
the first decay yields to the pair ground state, and so, to luminescence quenching;
the second, leaves \YbIII\ in its $^2F_{5/2}$ excited spin-orbit multiplet, which can ultimately lead to IVCT
luminescence.
We will see here how the magnitude of the energy barriers for these crossings
are responsible for either luminescence quenching, as in BaF$_2$, or interplay
between regular \dfem\  and IVCT luminescence, as in the CaF$_2$ and SrF$_2$ cases.
This is summarized in Fig.~\ref{FIG:Yb-cases}.

  \subsubsection{Luminescence quenching in BaF$_2$:\Ybtp}
  \label{SEC:BaF2}
 The crossing between \Ybiipiiip\ [$^2F_{7/2}5de_g,^2F_{7/2}$] and \Ybiiipiip\ [$^2F_{7/2},4f^{14}$]
 that enables quenching of the luminescence (step II$_{\rm qnch}$) 
 by direct non-raditive decay to the ground state
 after the first photon absorption (step I),
 i.e. from  \Ybiipiiip\ [1\Eu,1\Gsevu] to \Ybiiipiip\ [1\Gsevu,1\Aog],
 varies very strongly in the fluoride series
 (cf. Table~\ref{TAB:IVCTbarr} and Fig.~\ref{FIG:CaSrBaF2-YbIIYbIII-11-few}):
 1680~\cmm1\ (CaF$_2$), 818~\cmm1\ (SrF$_2$), 261~\cmm1\ in BaF$_2$.
 This result suggests that step II$_{\rm qnch}$ should result in
 appreaciable quenching of the luminescence in BaF$_2$ after \fdabs\  excitation.
 This observation holds both for IVCT and \dfem\  luminescence and 
 it is in line with experimental evidences.~\cite{MOINE:89}
 Then, we can say that BaF$_2$:\Ybtp-\YbIII\ belongs to the case $D$ depicted in Fig.~\ref{FIG:Yb-cases}.

  \subsubsection{Interplay between IVCT and \dfem\ emissions in CaF$_2$ and SrF$_2$}
  \label{SEC:CaF2SrF2}

The second important non-radiative pathway following excitation is 
step II (see Sec.~\ref{SEC:mechanism}), which enables IVCT non-radiative decay to the \Ybiiipiip\ [2\Gsevu,1\Aog]
excited state and is part of the IVCT luminescence mechanism. The shifts of the
energy barriers leading to step II from CaF$_2$ to SrF$_2$ allow to interpret the different
luminescent behaviour in these two hosts, as discussed next.

The energy barrier of IVCT crossing 
from \Ybiipiiip\ [1\Eu,1\Gsevu] 
to \Ybiiipiip\ [2\Gsevu,1\Aog]
increases from CaF$_2$ to SrF$_2$:
14~\cmm1\ (CaF$_2$), 148~\cmm1\ (SrF$_2$).
This suggest that [1\Eu,1\Gsevu] is  more stable in
SrF$_2$ than in CaF$_2$.
So, 
regular \YbII\ \dfem\
emission from this level should be more likely
in SrF$_2$, where, as a matter of fact, it has been observed.~\cite{MOINE:88}
For the same reason, its contribution to exciting the IVCT luminescence through steps II to IV
is much less likely in SrF$_2$ than in CaF$_2$, which correlates well with the
significantly overall smaller emission lifetime of the IVCT (anomalous) luminescence
measured in SrF$_2$ compared with CaF$_2$.~\cite{MOINE:89}
  Then, we can say that SrF$_2$:\Ybtp\ and CaF$_2$:\Ybtp\ belong, respectively,
  to a case intermediate between $C$ and $B$, and to case $B$ depicted in Fig.~\ref{FIG:Yb-cases}.

In contrast, the energy barriers for crossing from [1\Tou,1\Gsevu]
to [2\Gsevu,1\Aog]: 49~\cmm1\ (CaF$_2$) and 10~\cmm1\ (SrF$_2$), are comparable (although slightly smaller
in the latter), which suggests that 
[1\Tou,1\Gsevu] should contribute significantly to the excitation of the IVCT luminescence
in both hosts, especially at very low temperatures, since
non-radiative multiphonon decay from this level to the [1\Eu,1\Gsevu] should 
gradually quench this contribution in favour of a build up of the population of 
[1\Eu,1\Gsevu] on a temperature dependent basis.~\cite{GRIMM:05,PAN:08} 
Whereas this temperature dependent build up implies further excitation of the IVCT luminescence in CaF$_2$:\YbII
(because [1\Eu,1\Gsevu] preferetially undergoes step II of the IVCT luminesnce), 
it contributes to increasing  the intensity of the \YbII\ \dfem\  blue emission [1\Eu,1\Gsevu]$\rightarrow$[1\Aog,1\Gsevu] 
in SrF$_2$:\YbII\
(because \dfem\  emission is the preferred pathway for [1\Eu,1\Gsevu] depopulation in this material).
%
Altogether, these results allow to explain 
the different variation of the IVCT luminescence intensity with temperature in SrF$_2$ compared with CaF$_2$:
IVCT luminescence intensity reaches its maximum very sharply in SrF$_2$:\Ybtp, at very low temperature: 20~K,
and quenches already at 140~K; maximum intensity is reached more gradually at 110~K in CaF$_2$ and
quenching is observed at 180~K~(Ref.~\onlinecite{MOINE:89}) 
(this quenching has been connected with branching to the ground state in Sec.~\ref{SEC:mechanism}).
Correspondingly, the intensity of the \dfem\  emission [1\Eu,1\Gsevu]$\rightarrow$[1\Aog,1\Gsevu] 
in SrF$_2$ upon excitation in the [1\Tou,1\Gsevu] state (355~nm) 
is shown to be significantly higher than that measured
upon direct excitation in the [1\Eu,1\Gsevu] state (370~nm) beyond  140~K, which further reveals the
temperature dependent [1\Tou,1\Gsevu]$\rightarrow$[1\Eu,1\Gsevu] build up at the expense of 
the [1\Tou,1\Gsevu]$\rightarrow$[2\Gsevu,1\Aog] decay and IVCT luminescence.



Beyond step II, which has been discussed so far, the mechanism for IVCT
luminescence in SrF$_2$ is the same as in CaF$_2$ and leads to a very broad
band consisting of two overlapping bands corresponding to the three
\YbIII-\YbIII\ [1\Gsevu,2\Aou] $\rightarrow$ \YbII-\YbIII\ [$4f^{14}$,$^2F_{7/2}$] 
vertical transitions
at 17698, 17716, and 18121~\cmm1. Its FWHM is about 6000~\cmm1\, as it can be
seen in Fig.~\ref{FIG:CaSrF2-IVCTL}.
The peak energy and FWHM found experimentally are 12670~\cmm1\ and
4800~\cmm1. This indicates that the underestimation of the peak energy found in CaF$_2$ becomes
larger in the SrF$_2$ host.
In SrF$_2$,
the agreement between the calculated IR ESA spectrum and the one deduced from 
transient photoluminescence enhancement experiments in Ref.~\onlinecite{SENANAYAKE:13} 
is comparable to that in CaF$_2$. Both the theoretical and experimental spectra
can be compared in Table~\ref{TAB:CaSrF2-YbIIYbIII} and Fig.~\ref{FIG:CaSrF2-ESA}. 
The interpretation of the observed bands is the same as in the CaF$_2$ case.

  \subsubsection{IVCT absorption in \Ybtp-doped SrF$_2$}
  \label{SEC:SrF2}

Finally, we would like to report on the IVCT absorption bands which, according to our interpretation,
have been detected in the excitation spectra of \Ybtp-doped SrF$_2$ (Ref.~\onlinecite{MOINE:89})
even though they have not been identified as such.
We are not aware of similar experiments in the other hosts.

We will try to interpret the excitation spectra of the SrF$_2$:\YbII\ blue and red emissions~\cite{MOINE:89}
using the calculations presented in this paper.
According to them, 
a large number of close lying states of the \Ybiipiiip\ [$^2F_{7/2}5de_g$,$^2F_{7/2}$] 
configuration lie above its lowest state, [1\Eu,1\Gsevu], forming a dense manifold 5500~\cmm1\ wide.
This can be seen in 
Fig.~6 of Ref.~\onlinecite{SupMatRef}
(first manifold of states plotted with solid blue lines);
some of the levels are plotted in Fig.~\ref{FIG:CaSrBaF2-YbIIYbIII-11-few} and 
are tabulated in the absorption spectrum from the
[1\Aog,1\Gsevu] ground state in Table~\ref{TAB:CaSrF2-YbIIYbIII} (see also Ref.~\onlinecite{SupMatRef}).
In the vertical absorption spectrum from the [1\Aog,1\Gsevu] ground state (Table~\ref{TAB:CaSrF2-YbIIYbIII}),
this manifold is crossed by the 
stressed IVCT [$^2F_{5/2}$,$4f^{14}$]  branches  of \Ybiiipiip\ [2\Gsevu,1\Aog] and [2\Geigu,1\Aog] 
(dashed red lines in 
Fig.~6 of Ref.~\onlinecite{SupMatRef} and in
Fig.~\ref{FIG:CaSrBaF2-YbIIYbIII-11-few}) 
which are found at 28545 and 28809~\cmm1\ 
(4070 and 4333~\cmm1\ above the [1\Eu,1\Gsevu] state). 

Then, according to the mechanism discussed in Sec.~\ref{SEC:mechanism},
IVCT luminescence will be excited if these branches are reached 
either by crossings (like above) 
or by direct vertical IVCT absorptions from the ground state (like here, in the excitation spectrum). 
This explains the differences observed in the experimental excitation spectra:
The excitation spectra of the IVCT luminescence consists of one intense and very broad band
peaking at about 351~nm (28400~\cmm1)
as one can read in Fig.~3 of Ref.~\onlinecite{MOINE:89}.
This band can be assigned to the vertical IVCT absorptions
\Ybiipiiip\ [1\Aog,1\Gsevu]$\rightarrow$ \Ybiiipiip\ [2\Gsevu,1\Aog], [2\Geigu,1\Aog].
The band is broad due to the
large offset between the to minima [1\Aog,1\Gsevu] and [2\Gsevu,1\Aog] along the \Qet\ axis
(0.616~\AA). 
Interestingly, the broad band shows a shoulder in its high energy side, 
which indicates that 
non-radiative decay to the IVCT [2\Gsevu,1\Aog] branch from states lying above (like the electric dipole allowed [4\Tou,1\Gsevu])
could also excite the IVCT luminescence with sufficient efficiency
so as to be observed as a shoulder from the main IVCT absorption excitation channel.
Correspondingly, the excitation spectrum of the blue [1\Eu,1\Gsevu]$\rightarrow$[1\Aog,1\Gsevu]
emission shows the electronic structure of the 
[$^2F_{7/2}5de_g$,$^2F_{7/2}$] 
manifold interrupted by a dip, which coincides with the maximum of the intense IVCT absorption band
of the IVCT luminescence excitation spectrum. Also, the shoulder of the latter
and the fourth and highest intense peak of the former lie very close, which suggests the presence of 
the electric dipole allowed [1\Aog,1\Gsevu]$\rightarrow$[4\Tou,1\Gsevu] transition.
It is interesting to note that in most of the experiments reported for SrF$_2$:\Ybtp\
a larger wavelength (typically, 355~nm) has been used to study the
luminescent behaviour of the anomalous emission.

The IVCT absorption band we have just assigned can be interpreted as the following
one electron transfer:
\mbox{\Ybtp\ $4f_{5/2}$} $\rightarrow$ \mbox{\Ybthp\ $4f_{7/2}$}. 
It differs from the commonly observed IVCT absorptions in mixed valence compounds
in that the final state of the pair upon electron transfer is not the
stressed ground state, but, rather, a stressed excited state.
In this case, the \Ybthp\ part of the pair after electron transfer appears to be in the
$4f^{13}(^2F_{5/2})$ excited multiplet. This $4f$--$4f$ electron transfer is probably  weaker 
than the close lying $4f\rightarrow5d$ transitions in the absorption spectrum; however, it is
more efficient in exciting the IVCT luminescence, 
because of its direct decay (step II) not facing an energy barrier,
which explains its relative intensity in the excitation spectrum.

  \subsection{Interplay between IVCT and \dfem\ emissions in SrF$_2$ and SrCl$_2$}
  \label{SEC:SrF2SrCl2}

The dual character of the luminescence of Yb in SrF$_2$ we have just discussed, disappears in the SrCl$_2$ host, 
where the IVCT luminescence is not observed: only regular \dfem\ emission bands have been assigned in this  case. 
However, the emission spectrum of SrCl$_2$:\Ybtp\ is very complex.
This complexity stems precisely from the fact that the IVCT luminescence mechanism cannot occur, 
hence, \dfem\ bands which are not observable in SrF$_2$:\Ybtp, become uncovered in SrCl$_2$:\Ybtp.
The purpose of this
section is to explain why this is so even if the existence of 
\Ybiipiiip\ mixed valence pairs is likely.~\cite{PAN:08,SU:05:a,SU:05:b}
We do not intend to discuss
the complex \fdabs\ spectroscopy of SrCl$_2$:\Ybtp, which has been the subject of a number of  
experimental
and theoretical studies.~\cite{PIPER:67,WITZKE:73,PAN:08,SANCHEZ-SANZ:10:a,SANCHEZ-SANZ:10:c}

In the host series SrF$_2$, SrCl$_2$, it is the chemical change
 what leads the variations of the local properties
of the \Ybtp\ and \Ybthp\ moieties of the \Ybiipiiip\ pairs,
rather than the structural change.
In particular, the energies of the breathing mode of the donor 
\Dclus\ and acceptor \Aclus\ embedded clusters clearly show
the chemical change, so that the mean values of the ground state
vibrational frequencies
$\bar{\omega}_{a_{1g}} =$[\nue(1\Aog)+\nue(1\Gsevu)]/2, decrease from
402~\cmm1\ (0\%, SrF$_2$) to 246~\cmm1\ (39~\%, SrCl$_2$). 
These mean values basically determine the curvatures of the diabatic potential energy surfaces and 
of the 
IVCT energy diagrams of the
\Ybiipiiip\ embedded pairs. 
The change is very clear in 
Fig.~7 of Ref.~\onlinecite{SupMatRef}
and Fig.~\ref{FIG:SrF2Cl2-YbIIYbIII-11-few}.
As an effect,
the crossing points between the 
lowest \Ybiipiiip\ [$^2F_{7/2}5de_g$,$^2F_{7/2}$] states 
and the  stressed branches of the
\Ybiiipiip\ ground and excited configurations
[$^2F_{7/2}$,$4f^{14}$] and [$^2F_{5/2}$,$4f^{14}$]
are drastically shifted in SrCl$_2$
(see Fig.~\ref{FIG:SrF2Cl2-YbIIYbIII-11-few} and Table~\ref{TAB:IVCTbarr}).
And as a consequence, none of the
two non-radiative pathways is available in SrCl$_2$ anymore. 
  We can say that SrCl$_2$:\Ybtp-\YbIII\ belongs 
  to the case $A$ in Fig.~\ref{FIG:Yb-cases}.
Therefore, the first conclusion
driven from the SrCl$_2$:\Ybiipiiip\ IVCT energy diagrams is that the IVCT luminescence
observed in CaF$_2$ and SrF$_2$ cannot occur in this host. 
The second conclusion is that
not only is the  [1\Eu,1\Gsevu] state very stable, but also the
higher electric dipole allowed [1\Tou,1\Gsevu] state is, since it cannot decay through
step II, like in CaF$_2$ and SrF$_2$. This means that both states
can luminesce in a wide range of temperatures (blue arrows in Fig.~\ref{FIG:SrF2Cl2-YbIIYbIII-11-few})
and that the temperature dependent multiphonon relaxation
from [1\Tou,1\Gsevu] to [1\Eu,1\Gsevu], which influenced the intensities of the blue and red emissions
of SrF$_2$ discussed above, applies now to the interplay between radiative and non
radiative decays from [1\Tou,1\Gsevu], with the additional complexity arising from
the Boltzmann population of close lying states above the 1\Tou\ before room temperature is reached.
All of which results in a complex, but
well understood temperature dependence of the relative intensities and lifetimes of the 
two emission bands.~\cite{PAN:08}

In addition to the two
emission bands from [1\Tou,1\Gsevu] and [1\Eu,1\Gsevu], observed peaking at
26500 and 24700~\cmm1, respectively (this work: 25500 and 22900~\cmm1), Witzke \etal\ reported
other three emission bands peaking at 19000, 23900, and 25400~\cmm1, which were ruled out as
internal transitions of the \Ybtp\ ion and were called defect bands involving,
possibly, \Ybthp\ ions or \Ybiiipiip\ pairs.~\cite{WITZKE:73} 
The band peaking at 19000~\cmm1\
could be associated with the IVCT luminescence 
\Ybiipiiip\ [1\Eu,1\Gsevu]$\rightarrow$ \Ybiiipiip\ [1\Gsevu,1\Aog], calculated as a vertical transition at
17200~\cmm1\ (see green arrows in Fig.~\ref{FIG:SrF2Cl2-YbIIYbIII-11-few}).

 \section{\label{SEC:Conclusions}
        CONCLUSIONS}

  {\em Ab initio} quantum mechanical calculations of the electronic structure of \Ybiipiiip\ mixed valence pairs in fluorites 
  allow to conclude the existence of two-photon excited IVCT luminescence in Yb-doped CaF$_2$ and SrF$_2$.
  The IVCT emission is  found to be a bielectronic deexcitation involving 
  electron transfer from the donor to the acceptor moieties of the \Ybiipiiip\ pair,
  \mbox{\Ybtp\ $5de_g$ $\rightarrow$ \Ybthp\ $4f_{7/2}$},
  and   a \mbox{$4f_{7/2} \rightarrow 4f_{5/2}$} deexcitation within the \Ybtp\ $4f^{13}$ subshell:
  \YbII-\YbIII\ [\dFfh$5de_g$,\dFsh] $\rightarrow$ \YbIII-\YbII [\dFsh,$4f^{14}$].
Hence, it is a very slow rate radiative process. 
It is excited by a very efficient two-photon mechanism where each
photon provokes the same strong $4f^{14}$--1\Aog$\rightarrow$$4f^{13}(^2F_{7/2})5de_g$--1\Tou\ 
absorption in the \Ybtp\ part of the pair: the first one, from
the pair ground state;
the second one, from an excited state of the pair whose \Ybthp\ moiety is in the higher $4f^{13}(^2F_{5/2})$ 
spin-orbit multiplet.
The band widths of the emissions are very large, in analogy with the
wide band widths of well-known IVCT absorptions of transition metal mixed valence compounds.

The calculated IVCT energy diagrams show that two important non-radiative decay pathways may occur after
the first photon excitation, which involve non-radiative
\Ybiipiiip$\rightarrow$\Ybiiipiip\ electron transfer (see graphical conclusion Fig.~\ref{FIG:Yb-cases}).
One, leads to the ground
state [$4f^{13}(^2F_{7/2}),4f^{14}$] and quenches any emission (black line in Fig.~\ref{FIG:Yb-cases}); 
the other, leads to the excited state
[$4f^{13}(^2F_{5/2}),4f^{14}$] from where the (second-photon)  IVCT luminescence excitation
takes place (red line in Fig.~\ref{FIG:Yb-cases}). The structural change in the
CaF$_2$, SrF$_2$, BaF$_2$ series and the chemical change 
in the SrF$_2$, SrCl$_2$ series, influence the topology of the
calculated IVCT energy diagrams and shift the crossings with both decay pathways making
the quenching decay the most likely in BaF$_2$, the decay towards
IVCT luminescence excitation the most likely in CaF$_2$ and SrF$_2$, whereas none of the
two decays occur in the SrCl$_2$ case, all of which leads to total quenching
of any emission in BaF$_2$ (case D),
IVCT luminescence only in CaF$_2$ (case B), dual IVCT and $5d-4f$ emissions in SrF$_2$
(intermediate case between B and C), 
and only $5d-4f$ emissions in SrCl$_2$ (case A), as experimentally observed.

The electronic structure of the \Ybiipiiip\ pairs has been calculated within
the diabatic and independent \Ybtp\ and \Ybthp\ embedded cluster approximations.
The donor \Dclus\ and acceptor \Aclus\ \abinitio\ wavefunctions and energies have been
calculated using high quality, well-established \abinitio\ quantum chemical methods, including: extended
basis sets, non-dynamic and dynamic electron correlation, relativistic effects up to spin-orbit coupling,
and quantum mechanical host embedding, all of them, at the highest levels of methodology compatible
with the need to combine them all at once.
The two series of hosts chosen to demonstrate the capabilities and limitations
of the IVCT model presented
span a very complex luminescence scenario for validation.
In this context, the overall agreement of the theoretical and experimental results
including different types of samples (\Ybtp-doped, \Ybthp-doped) and
experimental techniques: emission, excitation, photoluminescence enhancement spectra and their
variation with temperature, is very satisfactory and allows to draw the
conclusion that the anomalous luminescence of \Ybtp\ associated so far with impurity-trapped excitons
is, rather, an  IVCT luminescence associated with \Ybiipiiip\ mixed valence pairs.
We also conclude that the broad band observed in the excitation spectrum of the so far called
anomalous emission of \Ybtp-doped SrF$_2$ is a broad IVCT absorption band 
corresponding to the following 
\Ybtp\ $4f_{5/2}$ $\rightarrow$ \Ybthp\ $4f_{7/2}$ electron transfer.
\acknowledgments
This work was partly supported by
a grant from Ministerio de Econom\'{\i}a y Competitividad, Spain
(Direcci\'on General de Investigaci\'on y Gesti\'on del Plan Nacional de I+D+i,
MAT2011-24586).

%

\begin{widetext}
 \iftoggle{usualpreprint}{
     \input{tables.tex}
 }{}
 \iftoggle{journal-like-bigfigures}{
 \clearpage
 \iftoggle{usualpreprint}{\renewcommand{\arraystretch}{0.64}}{} 
\setlength\LTcapwidth{0.95\textwidth}
\begin{longtable}{@{\extracolsep{\fill}}  r@{~}l cccc@{~~}r@{~~}lr@{~~}lr@{~~}lr@{~~}lr@{~~}l}
\caption{
          Spectroscopic constants and analyses of the 
          spin-orbit
          wave functions of the electronic states of independent  
          \Ybtp\ and \Ybthp\ -doped CaF$_2$ cubic defects. 
          Yb--F bond distances, $d_e$, in \AA; totally symmetric
          vibrational frequencies of the YbF$_8$ stretching mode, 
          \nue, in \cmm1;
          minimum-to-minimum energy differences, \Te,
          relative to the ground states, in \cmm1.
          1\Aog$\rightarrow i$\Tou\ absorption oscillator strengths of
          electric dipole allowed electronic transitions, $f_i$, are given relative
          to that to 1\Tou: $f_1$=2.013$\times$10$^{-2}$.
          Manifold averages and mean square deviations of the
          individual values are indicated.
          Higher lying excited states of \Ybtp\ can be found in Ref.~\onlinecite{SupMatRef}
        }
\label{TAB:Ca-SpinOrbit-short}

%
%
%
\\
\toprule
\\
\multicolumn{2}{c}{~State~} &
\multicolumn{1}{c}{~$d_e$~}   &
\multicolumn{1}{c}{~\nue~}  &
\multicolumn{1}{c}{~\Te~}   &
\multicolumn{1}{c}{$f_i/f_1$}   &
\multicolumn{8}{c}{weights of terms larger than 10\% $^{\rm a}$} \\
\cline{7-16}
\\
\multicolumn{14}{c}{
\Ybtp-doped CaF$_2$
} 
\\
\\ 
1&$A_{1g}$              & 2.329 & 417   &         0 & &
    100.0& 1$^1A_{1g}$                              
\\
\\
\multicolumn{14}{c}{
4$f^{13}$(\dFsh)$5de_g$
submanifold;
$\langle$$d_e$$\rangle$ = 2.315$\pm$0.001 ;
$\langle$\nue$\rangle$ = 423$\pm$1 
}
\\
\\
 1&$E_u$    & 2.317 & 423 & 23510 &      & 
     89.21&1~$^3T_{1u}$   & \\
 1&$T_{2u}$ & 2.317 & 423 & 23548 &      & 
     89.94&1~$^3T_{1u}$   & \\
 1&$T_{1u}$ & 2.316 & 423 & 25636 & 1.000& 
     42.33&1~$^3T_{1u}$   &
     34.07&1~$^1T_{1u}$   &
     19.47&1~$^3T_{2u}$   & \\
 1&$A_{2u}$ & 2.315 & 424 & 25704 &      & 
     99.57&1~$^3T_{2u}$   & \\
 2&$T_{2u}$ & 2.316 & 424 & 25730 &      & 
     68.39&1~$^3T_{2u}$   &
     30.16&1~$^1T_{2u}$   & \\
 2&$E_u$    & 2.315 & 422 & 26202 &      & 
     48.54&1~$^1E_{u}$    &
     25.35&1~$^3T_{2u}$   &
     25.02&2~$^3T_{1u}$   & \\
 3&$T_{2u}$ & 2.314 & 422 & 26420 &      & 
     48.54&1~$^3E_{u}$    &
     23.19&2~$^3T_{1u}$   &
     21.10&1~$^1T_{2u}$   & \\
 2&$T_{1u}$ & 2.314 & 423 & 26533 & 0.160& 
     50.49&1~$^3E_{u}$    &
     21.19&2~$^3T_{1u}$   &
     15.29&1~$^3T_{2u}$   & \\
 1&$A_{1u}$ & 2.315 & 424 & 26664 &      & 
     96.37&2~$^3T_{1u}$   & \\
 3&$T_{1u}$ & 2.316 & 421 & 27095 & 0.045& 
     55.67&2~$^3T_{1u}$   &
     26.78&2~$^1T_{1u}$   & \\
 3&$E_u$    & 2.315 & 422 & 28096 &      & 
     79.67&2~$^3T_{2u}$   &
     13.20&2~$^3T_{1u}$   & \\
 4&$T_{2u}$ & 2.315 & 423 & 28327 &      & 
     48.35&2~$^3T_{2u}$   &
     29.41&2~$^1T_{2u}$   & \\
 4&$T_{1u}$ & 2.316 & 421 & 28344 & 0.001& 
     56.72&2~$^3T_{2u}$   &
     20.97&2~$^1T_{2u}$   &
     11.40&2~$^1T_{1u}$   & \\
\\
\multicolumn{14}{c}{
4$f^{13}$(\dFfh)5$de_g$
submanifold;
$\langle$$d_e$$\rangle$ = 2.316$\pm$0.001 ;
$\langle$\nue$\rangle$ = 422$\pm$2
}
\\ 
\\
 2&$A_{1u}$ & 2.317 & 424 & 34004 &       &
     96.24&1~$^3T_{1u}$   & \\
 5&$T_{1u}$ & 2.316 & 423 & 35075 & 0.797 &
     51.63&1~$^3T_{1u}$   &
     21.10&1~$^1T_{1u}$   &
     16.08&1~$^3T_{2u}$   & \\
 4&$E_u$    & 2.315 & 423 & 35924 &       &
     64.80&1~$^3T_{2u}$   &
     18.78&1~$^1E_{u}$    & \\
 5&$T_{2u}$ & 2.315 & 423 & 36270 &       &
     43.72&1~$^1T_{2u}$   &
     22.91&1~$^3E_{u}$    &
     20.02&1~$^3T_{2u}$    & \\
 6&$T_{1u}$ & 2.316 & 421 & 36832 & 1.696 &
     41.99&1~$^3T_{2u}$   &
     36.54&1~$^1T_{1u}$   &
     15.64&1~$^3E_{u}$    & \\
 5&$E_u$    & 2.315 & 422 & 37291 &       &
     54.90&2~$^3T_{1u}$   &
     24.35&1~$^1E_{u}$    &
     19.89&2~$^3T_{2u}$   & \\
 6&$T_{2u}$ & 2.315 & 423 & 37344 &       &
     56.03&2~$^3T_{1u}$   &
     21.66&1~$^3E_{u}$    &
     14.64&2~$^1T_{2u}$   & \\
 7&$T_{1u}$ & 2.315 & 416 & 38161 & 0.016 &
     45.48&2~$^1T_{1u}$   &
     19.46&1~$^3E_{u}$    &
     16.83&2~$^3T_{1u}$   &
     16.23&2~$^3T_{2u}$   & \\
\\
\\
\multicolumn{14}{c}{
\Ybthp-doped CaF$_2$
} 
\\
\\
\multicolumn{14}{c}{
4$f^{13}$ manifold; 
$\langle$$d_e$$\rangle$ = 2.201$\pm$0.001 ;
$\langle$\nue$\rangle$ = 493$\pm$0
}
\\
\\
\multicolumn{14}{c}{
4$f^{13}$(\dFsh)
submanifold} \\
\\ 
1&$\Gamma_{7u}$ $^{\rm b}$         & 2.200 & 493 &     0 & &
     62.74&1~$^2A_{2u}$                              &
     37.26&1~$^2T_{2u}$                              & \\
1&$\Gamma_{6u}$    & 2.202 & 493 &   686 & &
     99.99&1~$^2T_{1u}$                              & \\
1&$\Gamma_{8u}$    & 2.202 & 494 &   729 & &
     63.93&1~$^2T_{2u}$                              &
     36.06&1~$^2T_{1u}$                              & \\
\\
\multicolumn{14}{c}{
4$f^{13}$(\dFfh)
submanifold} \\
\\ 
2&$\Gamma_{7u}$    & 2.201 & 494 & 10764 & &
     62.74&1~$^2T_{2u}$                              &
     37.26&1~$^2A_{2u}$                              & \\
2&$\Gamma_{8u}$    & 2.202 & 493 & 11196 & &
     63.94&1~$^2T_{1u}$                              &
     36.06&1~$^2T_{2u}$                              & \\
\\
\toprule
 \end{longtable}
\begin{minipage}{\mylongtablefootnote}
\begin{footnotesize}
$^{\rm a}$ 
Weights are given in \% and correspond to calculations at d(Yb-F) = 2.200~\AA\ and  2.383~\AA\
for \Ybthp\ and \Ybtp\ -doped CaF$_2$, respectively.

$^{\rm b}$ 
The \Ybtp\ 1\Aog$\rightarrow$\Ybthp\ 1$\Gamma_{7u}$ minimum-to-minimum energy difference is 6905~\cmm1.

\end{footnotesize}
\end{minipage}

 \clearpage
 \iftoggle{usualpreprint}{\renewcommand{\arraystretch}{0.64}}{}
\renewcommand{\thefootnote}{\alph{footnote}}
\setlength\LTcapwidth{0.85\textwidth}
\begin{longtable}{llcccccccc}
\caption{
         Absorption and emission peak positions of the \Ybiipiiip\ pair in CaF$_2$ and SrF$_2$
         calculated as total energy differences at the equilibrium geometries of
         the initial state.
         \dL\ and \dR\ are the Yb--F distances in the left and right YbF$_8$ moieties.
         Only data referred in the text are given. For all vertical transitions and
         data in BaF$_2$:\Ybiipiiip, see Ref.~\onlinecite{SupMatRef}.
         Experimental data are given in squared parentheses.
         Identification of calculated energy differences with absorption and emission transitions indicated in
         Figs.~\ref{FIG:CaF2-YbIIYbIII-11-few}  and \ref{FIG:CaSrBaF2-YbIIYbIII-11-few} are shown
         as roman numbers in parentheses. 
         Energies in \cmm1, distances in \AA.
         See text for details.
        }
\label{TAB:CaSrF2-YbIIYbIII}

\\
\toprule
\\
&&\multicolumn{6}{c}{\Ybiipiiip\ initial state    }\\
&& \multicolumn{2}{c}{[$4f^{14}$,\dFsh] }
&  \multicolumn{2}{c}{[$4f^{14}$,\dFfh] }
&  \multicolumn{2}{c}{[\dFfh$5de_g$,\dFsh]   } \\
&& \multicolumn{2}{c}{[1\Aog,1\Gsevu]              }
&  \multicolumn{2}{c}{[1\Aog,2\Gsevu]              }
&  \multicolumn{2}{c}{[2\Aou,1\Gsevu]              } \\
&& \multicolumn{2}{c}{\dL,\dR\                     }
&  \multicolumn{2}{c}{\dL,\dR\                     }
&  \multicolumn{2}{c}{\dL,\dR\                     } \\
&& CaF$_2$ & SrF$_2$ 
 & CaF$_2$ & SrF$_2$ 
 & CaF$_2$ & SrF$_2$ \\
&                  &  2.330,2.201  & 2.406,2.252                    
                   &  2.330,2.201  & 2.406,2.253                     
                   &  2.314,2.201  & 2.396,2.252      \\
\multicolumn{2}{c}{Final state} \\

\Ybiipiiip & [$4f^{14}$,\dFsh] \\
&&&&\multicolumn{2}{c}{\bf (III$_{em}$) \Ybthp\ luminescence} \\
              & $[$ 1\Aog,1\Gsevu $]$   &      0     &      0  &  -10762  & -10718   &  -33913  &  -34897        \\  
              & $[$ 1\Aog,1\Gsixu $]$   &    686     &    580  &  -10078  & -10140   &  -33226  &  -34316        \\  
              & $[$ 1\Aog,1\Geigu $]$   &    729     &    614  &  -10035  & -10106   &  -33183  &  -34283        \\  
\\
	      & [$4f^{14}$,\dFfh] \\
	      &                         & [10384]~$^{\rm a}$    &                                                           \\
              & $[$ 1\Aog,2\Gsevu $]$   &  10763     &  10719  &       0  &      0   &  -23149  &  -24177        \\  
	      &                         & [$\sim$10794]~$^{\rm a}$    &                                                           \\
              & $[$ 1\Aog,2\Geigu $]$   &  11196     &  11092  &          &          &  -22716  &  -23804        \\  
\\                                                                                                                              
\Ybiiipiip & [\dFsh,$4f^{14}$] \\
&&&&&&\multicolumn{2}{c}{\bf (V) IVCT luminescence}\\                                                                                                      
&&&&&& [-18200]~$^{\rm b}$ \\
&&&&&& [-18000]~$^{\rm c}$ &  [-13700]~$^{\rm c}$      \\
&&&&&& [-17400]~$^{\rm d}$ &  [-12000]~$^{\rm d}$      \\
              &  $[$ 1\Gsevu,1\Aog $]$   &  15238     &  17894  &          &          &  -20508  &  -18121        \\  
              &                          &            &         &          &          & [-17600]~$^{\rm b}$ &                \\
              &  $[$ 1\Gsixu,1\Aog $]$   &  15728     &  18290  &          &          &  -19999  &  -17716        \\  
              &  $[$ 1\Geigu,1\Aog $]$   &  15751     &  18307  &          &          &  -19974  &  -17698        \\  
\\                                                                                                           
\Ybiipiiip & [\dFsh$5de_g$,\dFsh] \\
&& [24814]~$^{\rm e}$    & [25316]~$^{\rm e}$ &          &    \\ 
              &  $[$ 1\Eu,1\Gsevu  $]$   &  23576     &  24475  &          &    \\ 
              &  $[$ 1\Ttu,1\Gsevu $]$   &  23614     &  24517  &          &    \\ 
              &  $[$ 1\Eu,1\Gsixu  $]$   &  24263     &  25055  &          &    \\ 
              &  $[$ 1\Ttu,1\Gsixu $]$   &  24301     &  25097  &          &    \\ 
              &  $[$ 1\Eu,1\Geigu  $]$   &  24305     &  25089  &          &    \\ 
              &  $[$ 1\Ttu,1\Geigu $]$   &  24344     &  25131  &          &    \\ 
&&\multicolumn{2}{c}{\bf (I) 1$^{\rm st}$ photon absorption}                                                  \\ %
&& [27400]~$^{\rm f}$    & [27950]~$^{\rm f}$ &          &    \\ 
              &  $[$ 1\Tou,1\Gsevu $]$   &  25706     &  26618  &          &    \\ 
&  others~$^{\rm g}$ \\ 
\\
\Ybiiipiip    & [\dFfh,$4f^{14}$] \\
              &                                       &            & {\bf IVCT absorption} \\
              &                                       &            & [$\sim$ 28490]~$^{\rm h}$ \\
              &  $[$ 2\Gsevu,1\Aog $]$   &  25922     &  28545  &          &    \\ 
              &  $[$ 2\Geigu,1\Aog $]$   &  26243     &  28809  &          &    \\ 
&  others~$^{\rm g}$ \\ 
\\
\Ybiipiiip & [\dF~$5de_g$,\dF] \\
&&&&&&\multicolumn{2}{c}{\bf IR ESA~$^{\rm i}$}\\
              &  $[$ 2\Aou,1\Gsevu $]$   &  34074     &  35003   &  23311   &  24284   &      0   &      0           \\
              &                          &            &          &          &          &   [250]  &   [178]          \\
              &  $[$ 1\Eu,2\Gsevu  $]$   &  34340     &  35194   &  23576   &  24475   &    270   &    195           \\
              &  $[$ 1\Ttu,2\Gsevu $]$   &  34378     &  35236   &  23614   &  24517   &    307   &    236           \\
              &                          &            &          &          &          & [650-950]& [500-670]        \\
              &  $[$ 2\Aou,1\Gsixu $]$   &  34761     &  35583   &  23996   &  24863   &    686   &    580           \\
              &  $[$ 1\Eu,2\Geigu  $]$   &  34772     &  35568   &  24007   &  24847   &    702   &    568           \\
              &  $[$ 2\Aou,1\Geigu $]$   &  34804     &  35617   &  24039   &  24896   &    729   &    614           \\
              &  $[$ 1\Ttu,2\Geigu $]$   &  34810     &  35610   &  24045   &  24889   &    739   &    609           \\
              &                          &            &          &          &          &  [1145]  &   [1273]         \\
              &  $[$ 5\Tou,1\Gsevu $]$   &  35145     &  36013   &  24382   &  25295   &   1069   &   1013           \\
              &  $[$ 5\Tou,1\Gsixu $]$   &  35832     &  36594   &  25067   &  25873   &   1756   &   1593           \\
 \Ybiiipiip\  &  $[$ 1\Gsevu,1\Eu  $]$   &            &          &          &          &   1775   &                  \\
 \Ybiipiiip\  &  $[$ 5\Tou,1\Geigu $]$   &  35874     &  36627   &  25109   &  25907   &   1799   &   1627           \\
 \Ybiiipiip\  &  $[$ 1\Gsevu,1\Ttu $]$   &            &          &          &          &   1809   &                  \\
 \Ybiipiiip\  &  $[$ 4\Eu,1\Gsevu  $]$   &  36005     &  36927   &  25242   &  26208   &   1917   &   1917           \\
              &  $[$ 5\Ttu,1\Gsevu $]$   &  36356     &  37301   &  25593   &  26582   &   2262   &   2292           \\
 \Ybiiipiip\  &  $[$ 1\Gsixu,1\Eu  $]$   &            &          &          &          &   2284   &                  \\
              &  $[$ 1\Geigu,1\Eu  $]$   &            &          &          &          &   2309   &                  \\
              &  $[$ 1\Gsixu,1\Ttu $]$   &            &          &          &          &   2319   &                  \\
              &  $[$ 1\Geigu,1\Ttu $]$   &            &          &          &          &   2343   &                  \\
&&&&\multicolumn{2}{c}{\bf (III) 2$^{\rm nd}$ photon absorption}\\                                                                       
 \Ybiipiiip\  &  $[$ 1\Tou,2\Gsevu $]$   &  36470     &  37337   &  25706   &  26618   &   2343   &   2336           \\
&  others~$^{\rm g}$ \\ 
\toprule
%
 \end{longtable}
\begin{minipage}{\mylongtablefootnote}
\begin{footnotesize}
$^{\rm a}$
Only the absorption line at 10384~\cmm1\ was assigned to a cubic site in Ref.~\onlinecite{KIRTON:67}; 
the splitting of the
$4f^{13}$(\dFfh) multiplet was expected to be similar to that found by Kiss (Ref.~\onlinecite{KISS:62})
for the isoelectronic ion CaF$_2$\Tmtp: 410~\cmm1.

$^{\rm b}$
From Ref.~\onlinecite{FEOFILOV:56} and \onlinecite{KAPLYANSKII:62}, below 200~K.

$^{\rm c}$
From Ref.~\onlinecite{REUT:76} at 77~K.

$^{\rm d}$
From Ref.~\onlinecite{MOINE:89} at 4.2~K.

$^{\rm e}$
Estimations from changes in the short time part of the fluorescence decay at 30~K, from Ref.~\onlinecite{MOINE:89}.

$^{\rm f}$
From absorption spectra from Ref.~\onlinecite{FEOFILOV:56} and  \onlinecite{KAPLYANSKII:62} at 20$^o$C and Ref.~\onlinecite{LOH:69} at 77~K.

$^{\rm g}$
Ref.~\onlinecite{SupMatRef}

$^{\rm h}$
Read from Fig. 2 of Ref.~\onlinecite{MOINE:89}; see text for details.

$^{\rm i}$
Experimental data from two frequency transient photoluminescence enhancement spectra at 10~K, 
from Ref.~\onlinecite{REID:11} (CaF$_2$) and Ref.~\onlinecite{SENANAYAKE:13} (SrF$_2$).

\end{footnotesize}
\end{minipage}

\clearpage
\begin{table}[h]
\iftoggle{usualpreprint}{\renewcommand{\arraystretch}{0.64}}{}
\caption{
         Calculated diabatic energy barriers for the \Ybiipiiip$\rightarrow$\Ybiiipiip\ electron transfer reaction
         within the \Ybiipiiip\ active pairs in 
         CaF$_2$, SrF$_2$, BaF$_2$, and SrCl$_2$ fluorite-type crystals. 
         [$Di,Aj$] and [$Ak,Dl$] are the electronic states of the pair before and after electron
         transfer. 
         The energy barrier for crossing in the forward, \mbox{[$Di,Aj$] $\rightarrow$ [$Ak,Dl$]},
         and backwards, \mbox{[$Di,Aj$] $\leftarrow$ [$Ak,Dl$]} reactions are given
         separated by a semicolon.
         The values of the Yb--X distances at the activated complex point, \dL,\dR, are given:
         they are, respectively, the Yb--X distances of the left ($L$) and right  ($R$) YbX$_8$  moieties.
         Energy barriers in bold format can be seen in Figures 
         \ref{FIG:CaF2-YbIIYbIII-11-few},
         \ref{FIG:CaSrBaF2-YbIIYbIII-11-few}, and
         \ref{FIG:SrF2Cl2-YbIIYbIII-11-few},
         See text for details.
        }
\label{TAB:IVCTbarr}

\begin{ruledtabular}
\begin{tabular}{crlccccc}
\\
\\
 &&& \multicolumn{4}{c}{forward; backwards diabatic energy barriers for IVCT} \\
   & \multicolumn{1}{r}{(YbX$_8$)$^{6-}_L$ + (YbX$_8$)$^{5-}_R$  $\rightarrow$ } 
   & \multicolumn{1}{l}{(YbX$_8$)$^{5-}_L$ + (YbX$_8$)$^{6-}_R$}            
   & \multicolumn{4}{c}{\dL,\dR\ at the crossing point                   } \\
\\
   & \multicolumn{1}{r}{ \Ybiipiiip\ $\rightarrow$     } & \multicolumn{1}{l}{\Ybiiipiip} 
   &   CaF$_2$ 
   &   SrF$_2$
   &   BaF$_2$
   &   SrCl$_2$   \\
\\
                   & [$Di$,$Aj$]                  $\rightarrow$     & [$Ak,Dl$] \\
\\                 
                   & [$4f^{14}$,$4f^{13}(7/2)$]   $\rightarrow$         & [$4f^{13}(5/2),4f^{14}$]       \\
\\
                   & [1\Aog,1\Gsevu]              $\rightarrow$         & [2\Gsevu,1\Aog]                    
                                         & 11048; {\bf 285}    & 11581; {\bf 862}   & 12390; {\bf 1700}  & 10837; {\bf 196}  \\
       &&                                & 2.220,2.311   & 2.285,2.377  & 2.349,2.442  & 2.711,2.878 \\
\\ 
                   & [$4f^{13}(7/2)5de_g$,$4f^{13}(7/2)$] $\rightarrow$        & [$4f^{13}(5/2),4f^{14}$]       \\
\\
&&&\multicolumn{2}{c}{\bf (II) crossing} \\
                   & [1\Eu ,1\Gsevu]        $\rightarrow$         & [2\Gsevu,1\Aog]                     
                                         & {\bf 14};   12741   & {\bf 148}; 13874   & {\bf 456}; 15281   & {\bf 1826}; 14375 \\
       &&                                & 2.312,2.205   & 2.383,2.266  & 2.448,2.322  & 2.880,2.671 \\
                   & [1\Ttu ,1\Gsevu]              $\rightarrow$        & [2\Gsevu,1\Aog]                     
                                         & 10;   12775   & 142; 13910   & 453; 15323   & 1847; 14433 \\
       &&                                & 2.313,2.205   & 2.380,2.263  & 2.446,2.320  & 2.882,2.672 \\
                   & [1\Tou ,1\Gsevu]              $\rightarrow$        & [2\Gsevu,1\Aog]                     
                                         & 49   ; 14903  & 10;   15877  & 166; 17164   & 3400; 17963 \\
       &&                                & 2.323,2.197   & 2.391,2.255  & 2.460,2.314  & 2.903,2.654 \\
\\
                   & [$4f^{13}(7/2)5de_g$,$4f^{13}(7/2)$] $\rightarrow$        & [$4f^{13}(7/2),4f^{14}$]       \\
\\
&&&&&\multicolumn{1}{c}{\bf (II$qnch$) quenching} \\
                   & [1\Eu ,1\Gsevu]              $\rightarrow$         & [1\Gsevu,1\Aog]                                    
                                         & {\bf 1680}; 25174   & {\bf 818}; 25263   & {\bf 261}; 25775   &  9416; 32606 \\
       &&                                & 2.360,2.159   & 2.435,2.222  & 2.501,2.278  & 2.978,2.612 \\
                   & [1\Ttu ,1\Gsevu]              $\rightarrow$        & [1\Gsevu,1\Aog]                    
                                         & 1701; 25229   & 819; 25306   & 266; 25825   & 9425; 32652 \\
       &&                                & 2.364,2.162   & 2.435,2.222  & 2.504,2.280  & 2.978,2.612 \\
                   & [1\Tou ,1\Gsevu]              $\rightarrow$        & [1\Gsevu,1\Aog]                    
                                         & 2561; 28179   & 1389; 27976  & 570; 28256   & 12846; 38050 \\
       &&                                & 2.375,2.154   & 2.444,2.212  & 2.510,2.267  & 3.025,2.612 \\
\\
        		   & [$4f^{13}5de_g$,$4f^{13}$] $\rightarrow$           & [$4f^{13}(7/2),4f^{13}(7/2)5de_g$]       \\
\\
                   & [2\Aou,1\Gsevu]              $\rightarrow$         & [1\Gsevu,1\Eu]                     
                                         & {\bf 76};   10570   & {\bf 385}; 10909   & {\bf 912}; 11470   &  2838; 13393 \\
       &&                                & 2.306,2.209   & 2.367,2.268  & 2.433,2.331  & 2.914,2.675 \\
                   & [1\Eu  ,2\Gsevu]              $\rightarrow$        & [1\Gsevu,1\Eu ]                    
                                         & 55.5; 10818   & 347; 11067   & 878; 11567   & 2853; 13494 \\
       &&                                & 2.308,2.209   & 2.369,2.268  & 2.434,2.331  & 2.916,2.676 \\
\\
                           & [$4f^{13}5de_g$,$4f^{13}$] $\rightarrow$           & [$4f^{13}(5/2),4f^{14}$]       \\
\\
                   & [2\Aou,1\Gsevu]              $\rightarrow$         & [2\Gsevu,1\Aog]                    
                                         & {\bf 1650}; 24868   & {\bf 800}; 25051   & {\bf 262}; 25646   & 9491; 32595 \\
       &&                                & 2.360,2.160   & 2.431,2.220  & 2.503,2.280  & 2.978,2.612 \\
                   & [1\Eu  ,2\Gsevu]              $\rightarrow$        & [2\Gsevu,1\Aog ]                    
                                         & 1774; 25264   & 861; 25307   & 277; 25792   & 9522; 32712 \\
       &&                                & 2.365,2.162   & 2.436,2.222  & 2.564,2.241  & 2.979,2.612 \\
\\

\end{tabular}
\end{ruledtabular}
\end{table}

 }{}
%
\iftoggle{journal-like-bigfigures}{
    \def\escalafig{1.0}\clearpage
    \begin{figure}[ht]
\resizebox{\escalafig\textwidth}{!}{
  \includegraphics{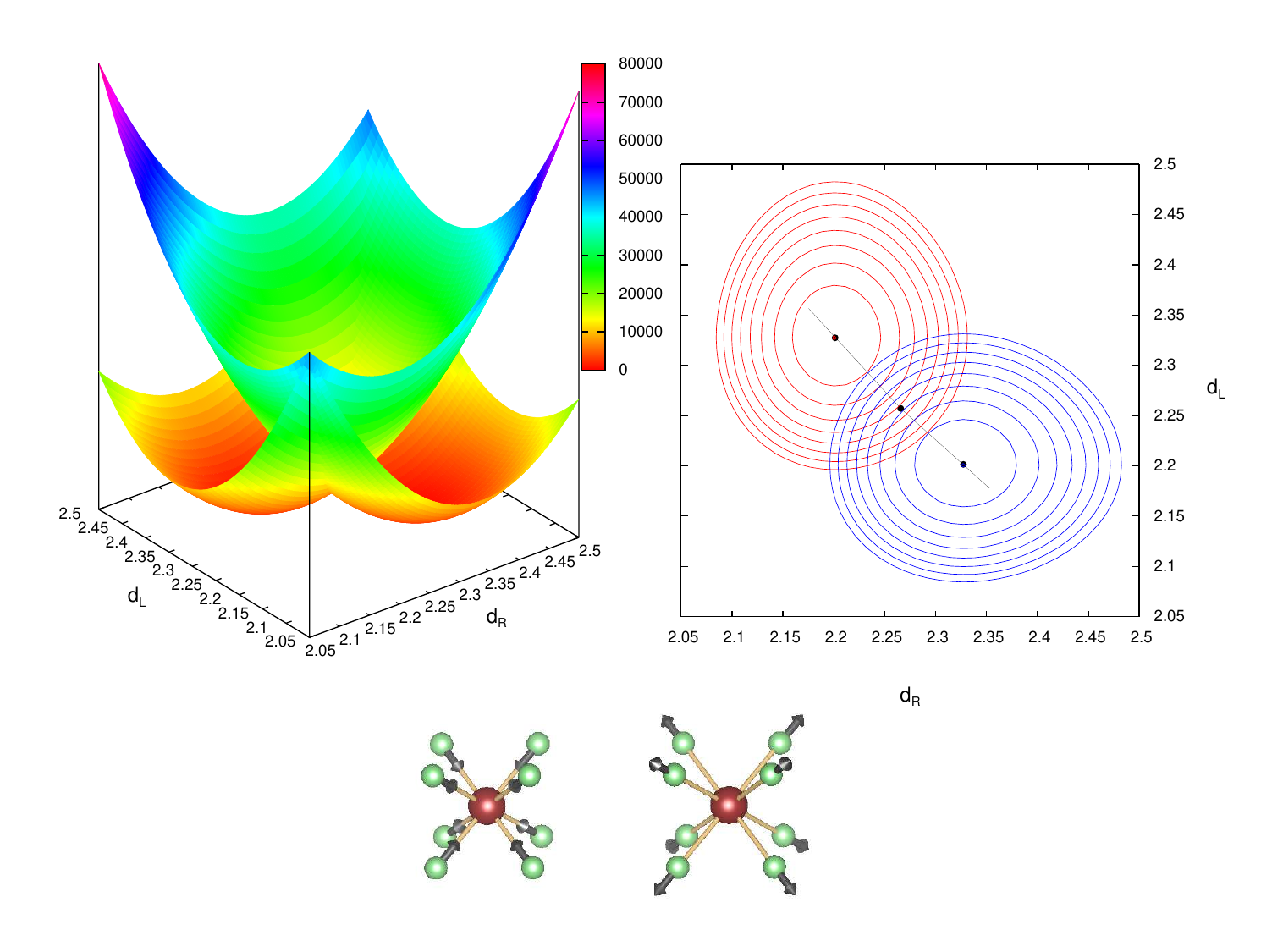}
}
\caption{
         Results of quantum mechanical calculations on Yb-doped CaF$_2$ crystals.
         Top graphs:
         The two symmetric diabatic potential energy surfaces corresponding to
         the [1\Aog,1\Gsevu]
         ground state of the \Ybiipiiip\ pair before 
         (YbF$_8$)$^{6-}_L$--(YbF$_8$)$^{5-}_R$ [1\Aog,1\Gsevu] (red in right graph) and after
         (YbF$_8$)$^{5-}_L$--(YbF$_6$)$^{5-}_R$  [1\Gsevu,1\Aog] (blue in right graph) electron transfer
         are plotted (left graph) and projected (right graph)
         in the \dL,\dR\ plane.
         The Yb--F distances of the left (\dL) and
         right (\dR) moieties of the pair are given in \AA.
         Top left graph:
         Energies are given in ~\cmm1\ and are referred to the energy of the
         two equivalent ground state minima. 
         Top right graph:
         The minimal energy reaction path connecting the two equivalent minima
         through the activated complex is indicated.
         Bottom graph:
         Displacements of the X atoms of the (YbX$_8$)$_L$--(YbX$_8$)$_R$
         moieties along the electron transfer reaction coordinate $Q_{et}$.
         See text for details.
        }
\label{FIG:ivct-gs-Qet}
    \end{figure}
    \def\escalafig{0.9}\clearpage 
    \begin{figure}[ht]
\resizebox{\escalafig\textwidth}{!}{
  \includegraphics{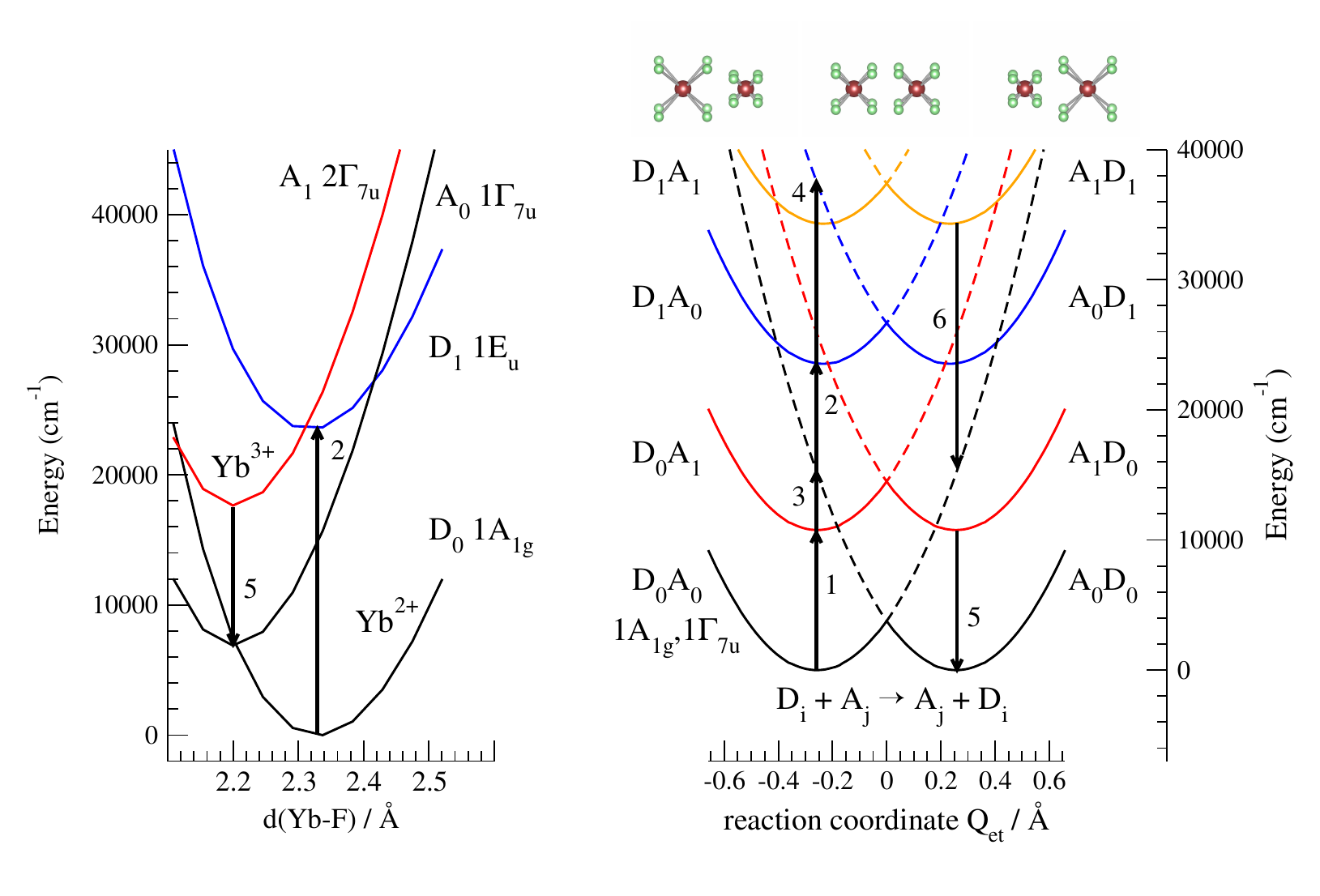}
}

\caption{
         Results of quantum mechanical calculations on Yb-doped CaF$_2$ crystals.
         Left graph:
         potential energy curves of the ground and one excited 
         state of the independent (YbF$_8$)$^{6-}$ and (YbF$_8$)$^{5-}$ embedded clusters.
         Right graph: 
         Diabatic IVCT energy diagram for \Ybiipiiip\ embedded pairs.
         The symmetric
         (YbF$_8$)$^{6-}_L$--(YbF$_8$)$^{5-}_R$ $D_iA_j$ and 
         (YbF$_8$)$^{5-}_L$--(YbF$_6$)$^{5-}_R$ $A_jD_i$ branches 
         are plotted $vs.$ the normal electron transfer reaction coordinate $Q_{et}$ of
         the ground state.
         $D_i$ and $A_j$ refer to the parent independent embedded cluster states.
         Top right graph: Qualitative representations of the structure of the left and right
         moieties of the pair for \Qet$<$0, =0 (activated complex), and $>0$
         are plotted.
         See text for details.
        }
\label{FIG:monomers-dimer}
    \end{figure}
    \def\escalafig{1.0}\clearpage 
    \begin{figure}[ht]
\resizebox{\escalafig\textwidth}{!}{
  \includegraphics{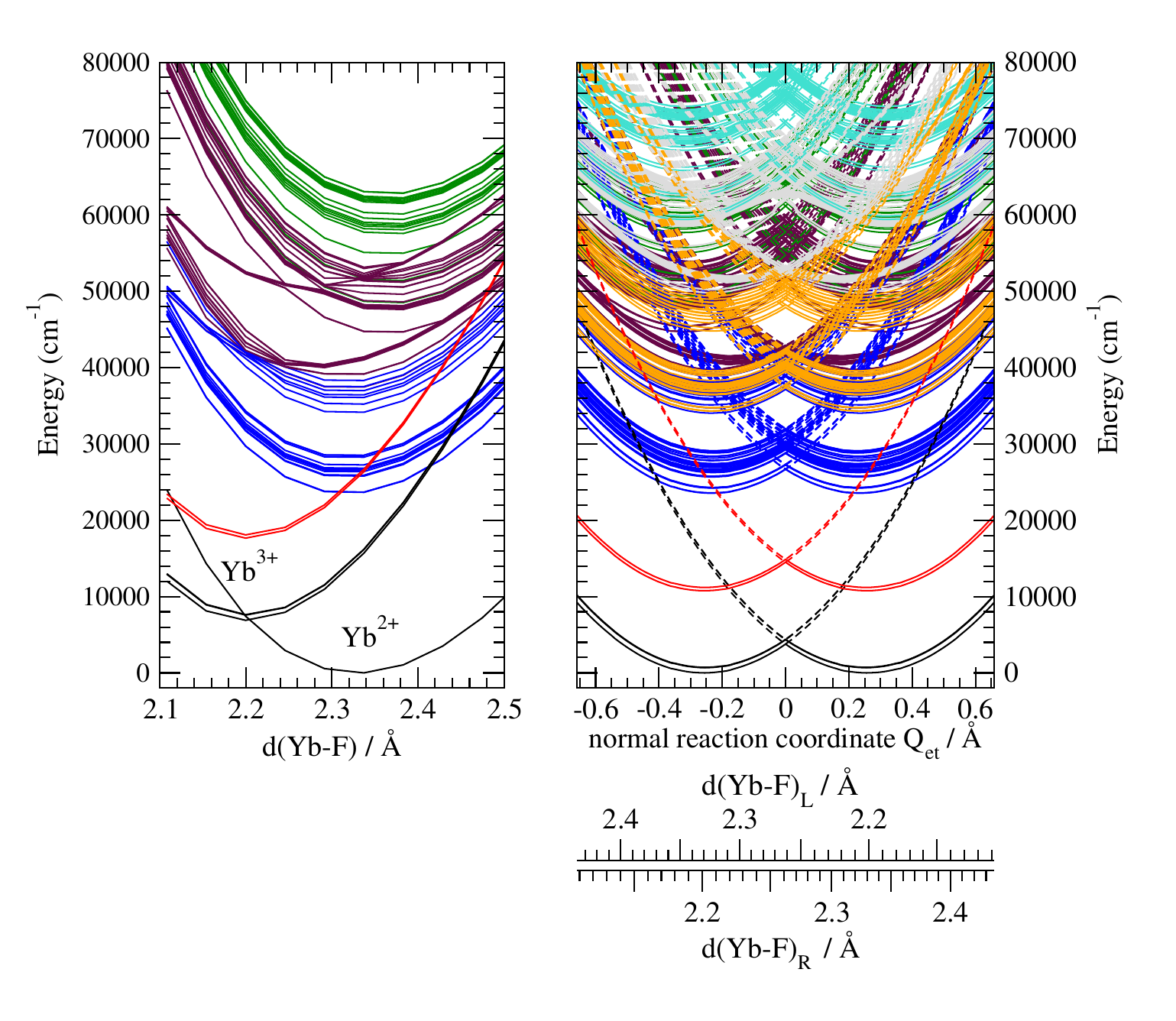}
}
\caption{
         Results of quantum mechanical calculations on Yb-doped CaF$_2$ crystals.
         Left graph:
         potential energy curves of the ground and excited 
         states of independent donor (YbF$_8$)$^{6-}$ and acceptor 
         (YbF$_8$)$^{5-}$ embedded clusters.
         Right graph: 
         Diabatic IVCT energy diagram for \Ybiipiiip\ embedded pairs.
         The symmetric
         (YbF$_8$)$^{6-}_L$--(YbF$_8$)$^{5-}_R$ and 
         (YbF$_8$)$^{5-}_L$--(YbF$_6$)$^{5-}_R$  branches 
         of the ground and excited states of \Ybiipiiip\ pairs
         are plotted $vs.$ the normal electron transfer reaction coordinate $Q_{et}$
         for the pair ground state;
         the values of the Yb--F distance of the left and right YbF$_8$ moieties, 
         \dL\ and \dR\ are indicated.
         Colors in left graph: black  4$f^{14}$ and 4$f^{13}$(7/2);
         red:  4$f^{13}$(5/2);
         blue: 4$f^{13}$5$de_g$;
         marroon: interacting 4$f^{13}$5$d$ and $4f^{13}a_{1g}^{YbTE}$;
         green: 4$f^{13}$5$dt_{2g}$ manifolds.
         Colors in right graph [\Ybtp,\Ybthp]:
         black   [4$f^{14}$,4$f^{13}$(7/2)],
         blue    [4$f^{13}$5$de_g$,4$f^{13}$(7/2)],
         marroon [interacting 4$f^{13}$5$d$ and $4f^{13}a_{1g}^{YbTE}$,4$f^{13}$(7/2)],
         green   [4$f^{13}$5$dt_{2g}$,4$f^{13}$(7/2)];
         the four previous colours become 
         red, orange, grey, and turquoise, when the
         \Ybtp\ states are combined with the \Ybthp\ 4$f^{13}$(5/2)
         states, instead.
        }
\label{FIG:CaF2-monomers-dimer}
    \end{figure}
    \def\escalafig{0.6}\clearpage 
    \begin{figure}[ht]
\resizebox{\escalafig\textwidth}{!}{
  \includegraphics{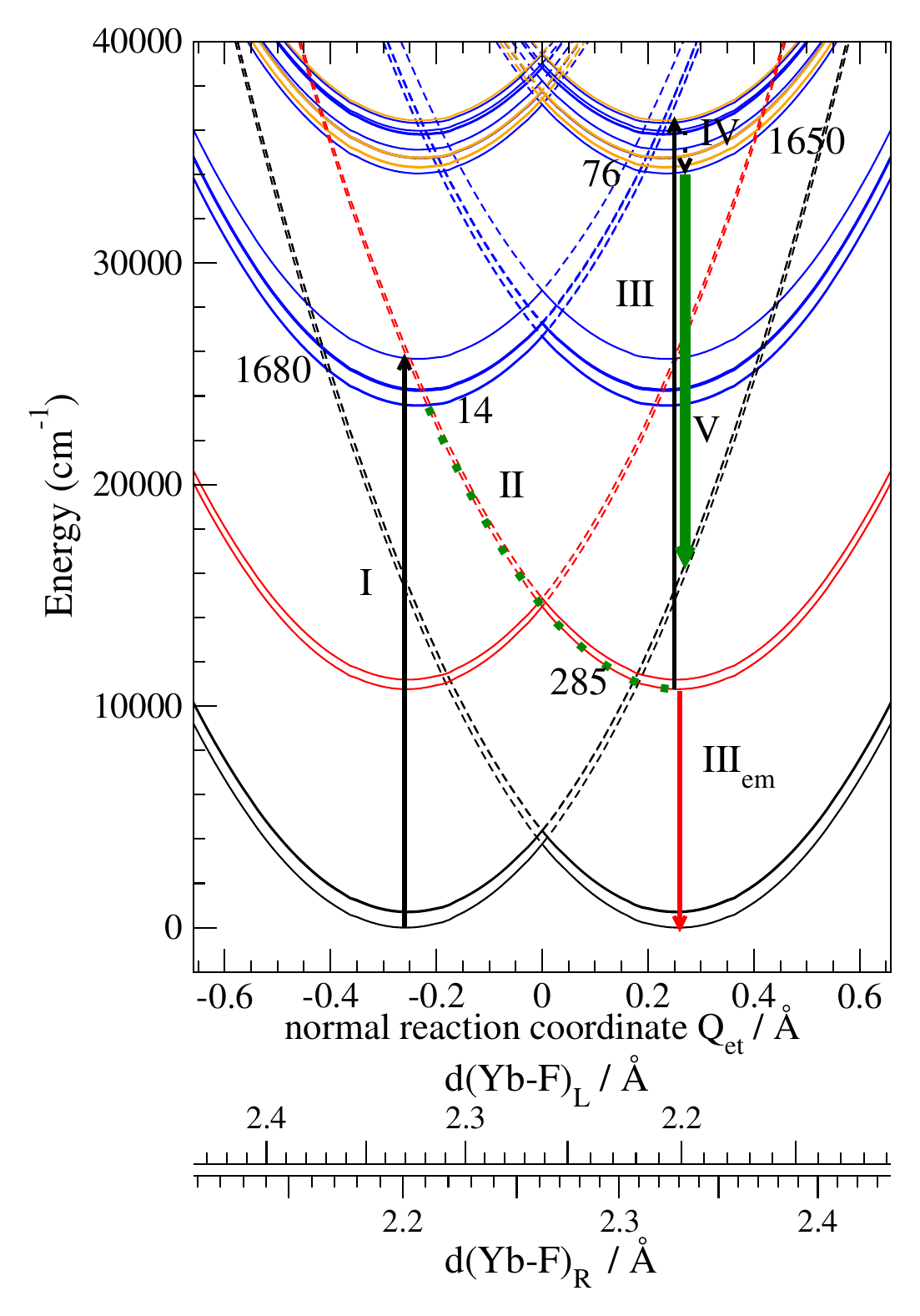}
}
\caption{
         Results of quantum mechanical calculations of the diabatic IVCT energy diagram for
         \Ybiipiiip\ pairs in Yb-doped CaF$_2$ crystals along the ground state normal
         electron transfer reaction coordinate Q$_{et}$.
         Mechanism of the IVCT luminescence
         of Yb-doped CaF$_2$ crystals: steps I to V.
         Mechanism of the excitation of the IR luminescence of \Ybthp\
         through the lowest $4f-5d$ absorption band of \Ybtp: steps I, II, III$_{em}$.
         Energy barriers in \cmm1\ are indicated next to the crossing points between
         two electronic states of the pairs; see details in Table~\ref{TAB:IVCTbarr}.
         See caption of Fig.~\ref{FIG:CaF2-monomers-dimer} and text for details.
        }
\label{FIG:CaF2-YbIIYbIII-11-few}

    \end{figure}
    \def\escalafig{0.7}\clearpage 
    \begin{figure}[ht]
\resizebox{\escalafig\textwidth}{!}{
  \includegraphics{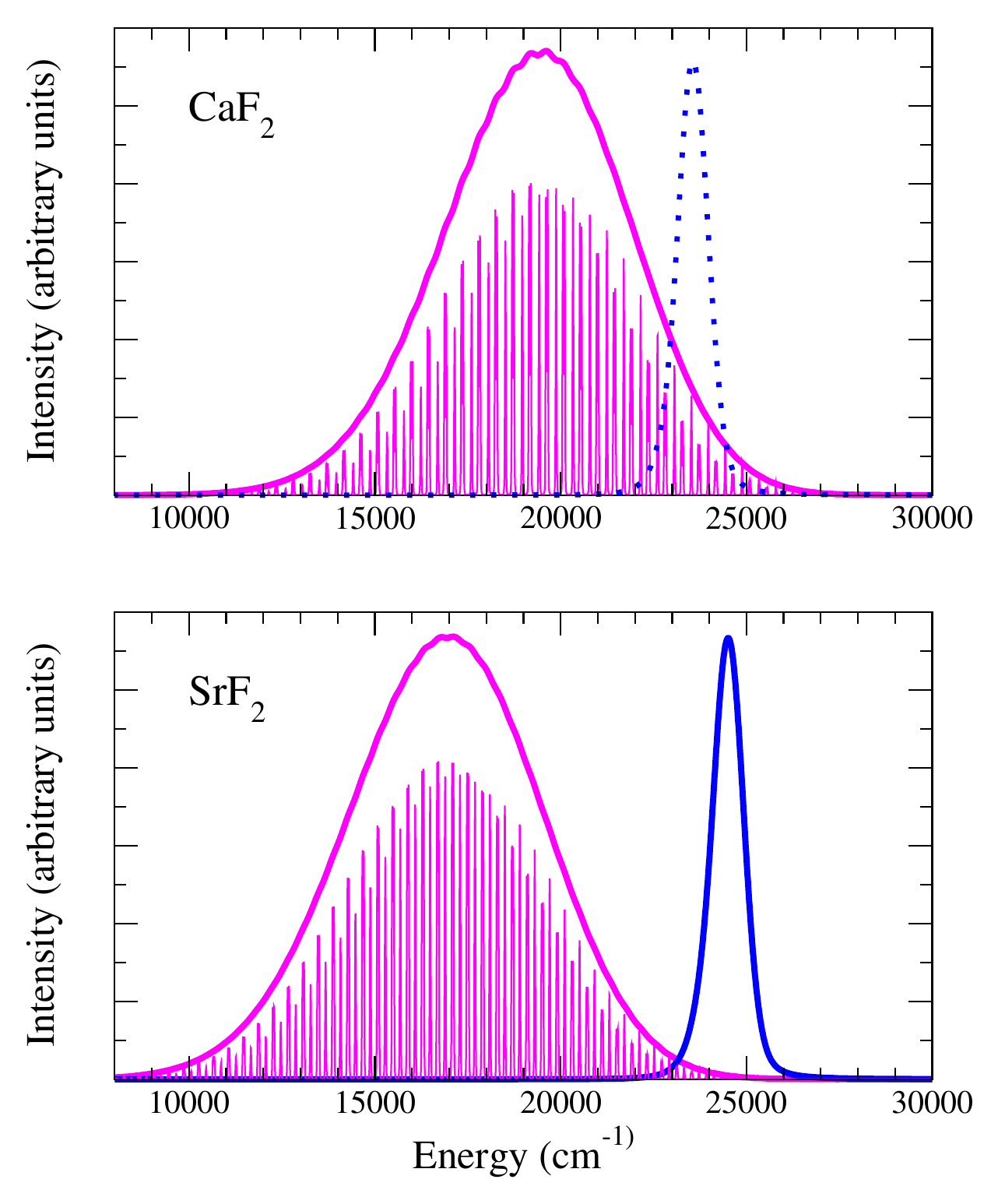}
}
\caption{
         Calculated band profiles of the 
         \Ybiipiiip\ [2\Aou,1$\Gamma_{7u}$] $\rightarrow$ \Ybiiipiip\ [1$\Gamma_{8u,6u,7u}$,1\Aog]
         intervalence charge transfer luminescence of
         (YbF$_8$)$^{6-}$--(YbF$_8$)$^{5-}$ embedded cluster pairs (magenta) and
         of the 1$E_u$ $\rightarrow$ 1\Aog\  emission band 
         of embedded (YbF$_8$)$^{6-}$ (blue) in CaF$_2$ and Sr$_2$. 
         Dotted line: not observable emission.
         Arbitrary values of the oscillator strengths have been used.
         See text for details.
        }
\label{FIG:CaSrF2-IVCTL}
    \end{figure}
    \def\escalafig{0.8}\clearpage
    \begin{figure}[ht]
\resizebox{\escalafig\textwidth}{!}{
  \includegraphics{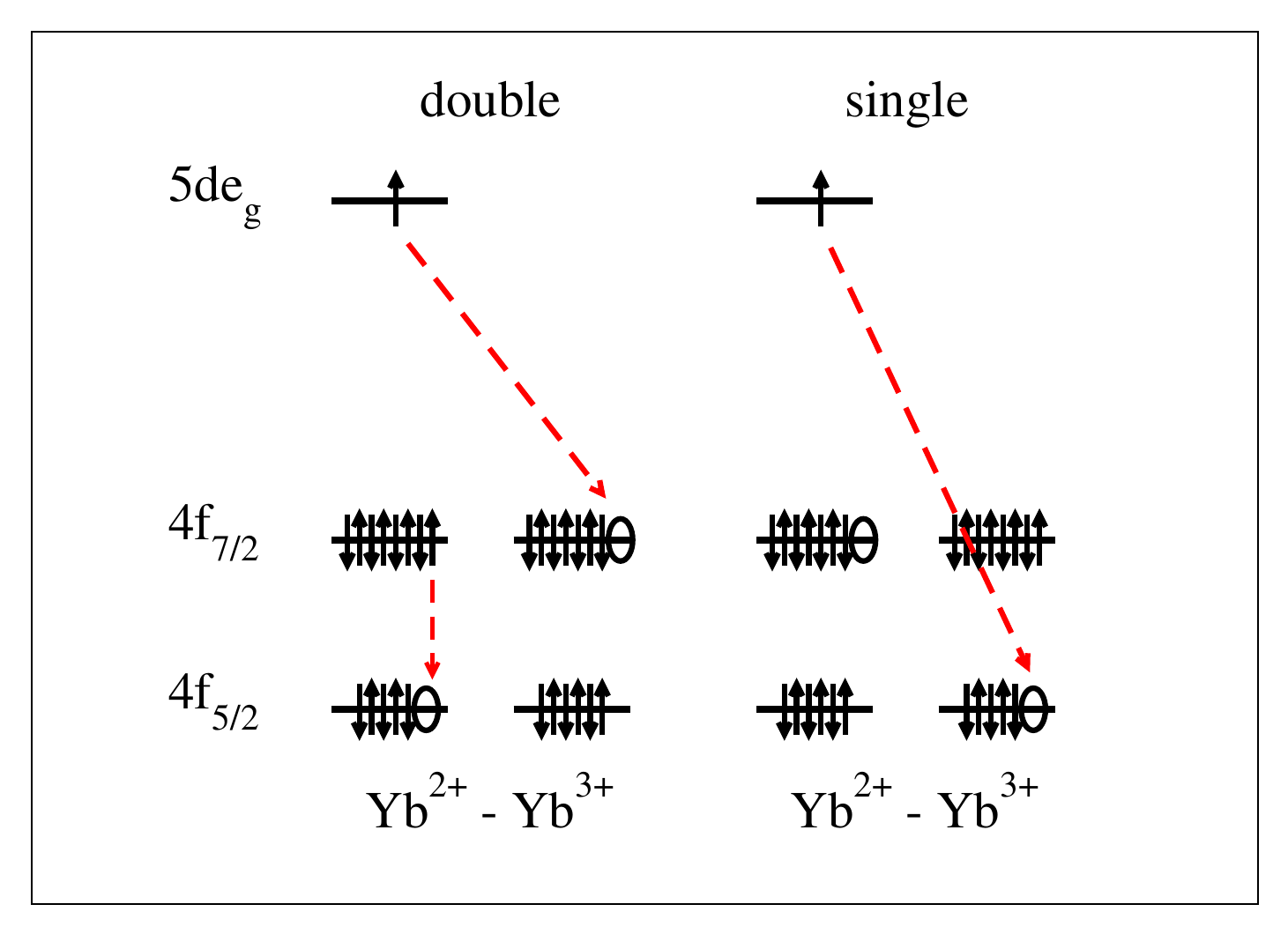}
}
\caption{
         Schematic representation of slow and fast IVCT emissions corresponding to
         a double deexcitation (double) consisting of an
         \mbox{\Ybtp\ $5de_g$ $\rightarrow$ \Ybthp\ $4f_{7/2}$} electron transfer
         accompanied by a $4f_{7/2} \rightarrow 4f_{5/2}$ deexcitation within the
         $4f^{13}$ subshell of \Ybtp;
         and a single deexcitation (single) consisting of an
         \mbox{\Ybtp\ $5de_g$ $\rightarrow$ \Ybthp\ $4f_{5/2}$} electron transfer.
        }
\label{FIG:ivct-scheme}
    \end{figure}
    \def\escalafig{0.7}\clearpage
    \begin{figure}[ht]
\resizebox{\escalafig\textwidth}{!}{
  \includegraphics{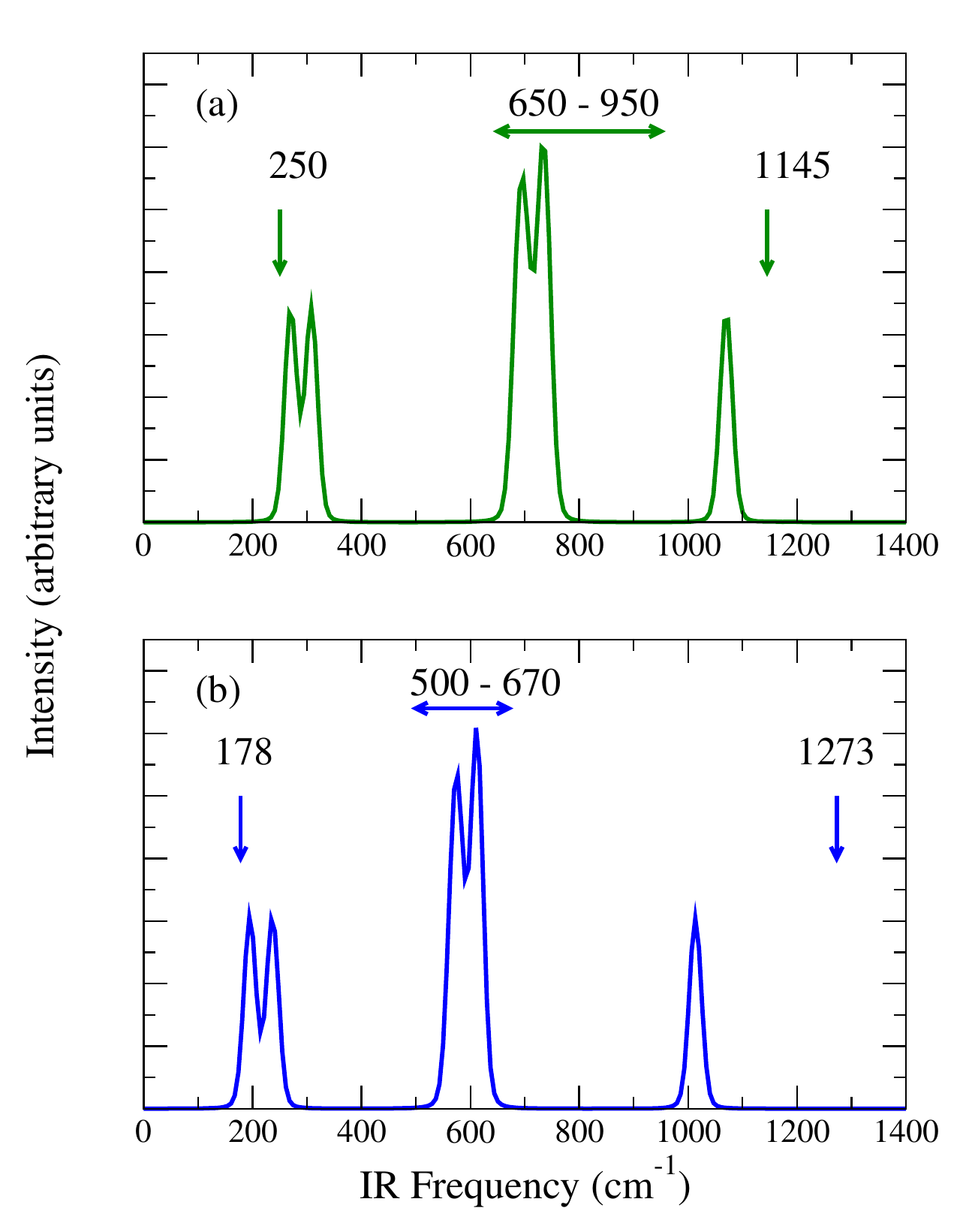}
}
\caption{
         Calculated excited state absorption spectrum of Yb--doped CaF$_2$ (a)
         and SrF$_2$ (b)
         originating in the \Ybiipiiip\ [2\Aou,1$\Gamma_{7u}$] excited state of the
         (YbF$_8$)$^{6-}$--(YbF$_8$)$^{5-}$ embedded cluster pairs.
         Transition energies are taken from Table~\ref{TAB:CaSrF2-YbIIYbIII}. All transitions
         are arbitrarily assigned the same oscillator strength value.
         Experimental values from Ref.~\onlinecite{REID:11} (a) and
         \onlinecite{SENANAYAKE:13} (b) are indicated with arrows.
        }
\label{FIG:CaSrF2-ESA}
    \end{figure}
    \def\escalafig{0.9}\clearpage
    \begin{figure}[ht]
\resizebox{\escalafig\textwidth}{!}{
  \includegraphics{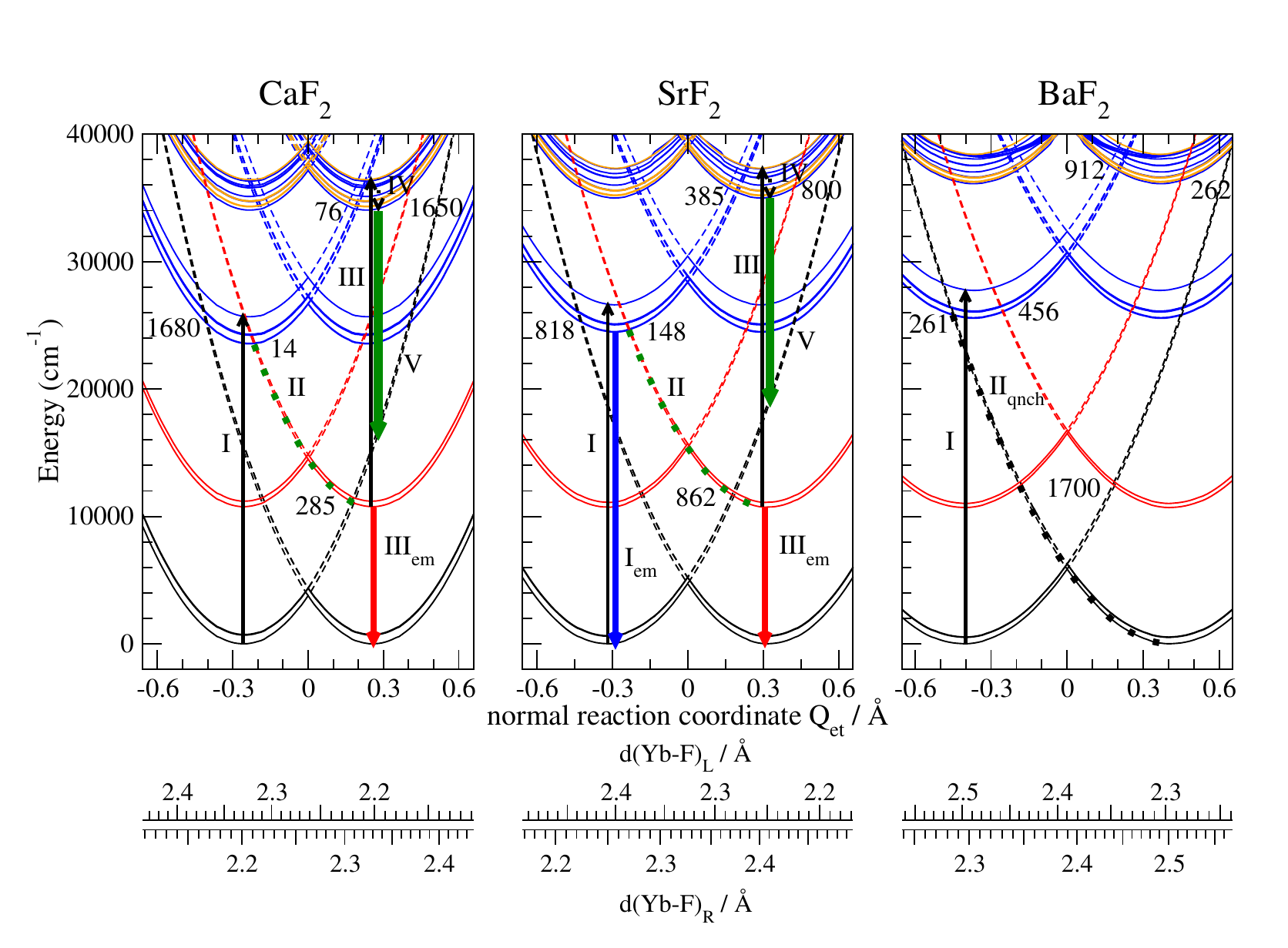}
}
\caption{
         Results of quantum mechanical calculations of the diabatic IVCT energy diagram for
         \Ybiipiiip\ pairs in Yb-doped CaF$_2$, SrF$_2$, and BaF$_2$ 
         crystals along the ground state normal
         electron transfer reaction coordinate Q$_{et}$.
         CaF$_2$ and SrF$_2$:
         IVCT luminescence: steps I to V;
         excitation of the IR luminescence of \Ybthp: steps I, II, III$_{\rm em}$.
         BaF$_2$:
         quenching of the Yb luminescence: Steps I and II$_{\rm qnch}$.
         Energy barriers in \cmm1\ are indicated next to the crossing points between
         two electronic states of the pairs; see details in Table~\ref{TAB:IVCTbarr}.
         See caption of Fig.~\ref{FIG:CaF2-monomers-dimer} and text for details.
        }
\label{FIG:CaSrBaF2-YbIIYbIII-11-few}
    \end{figure}
    \def\escalafig{1.0}\clearpage
    \begin{figure}[ht]
\resizebox{\escalafig\textwidth}{!}{
  \includegraphics{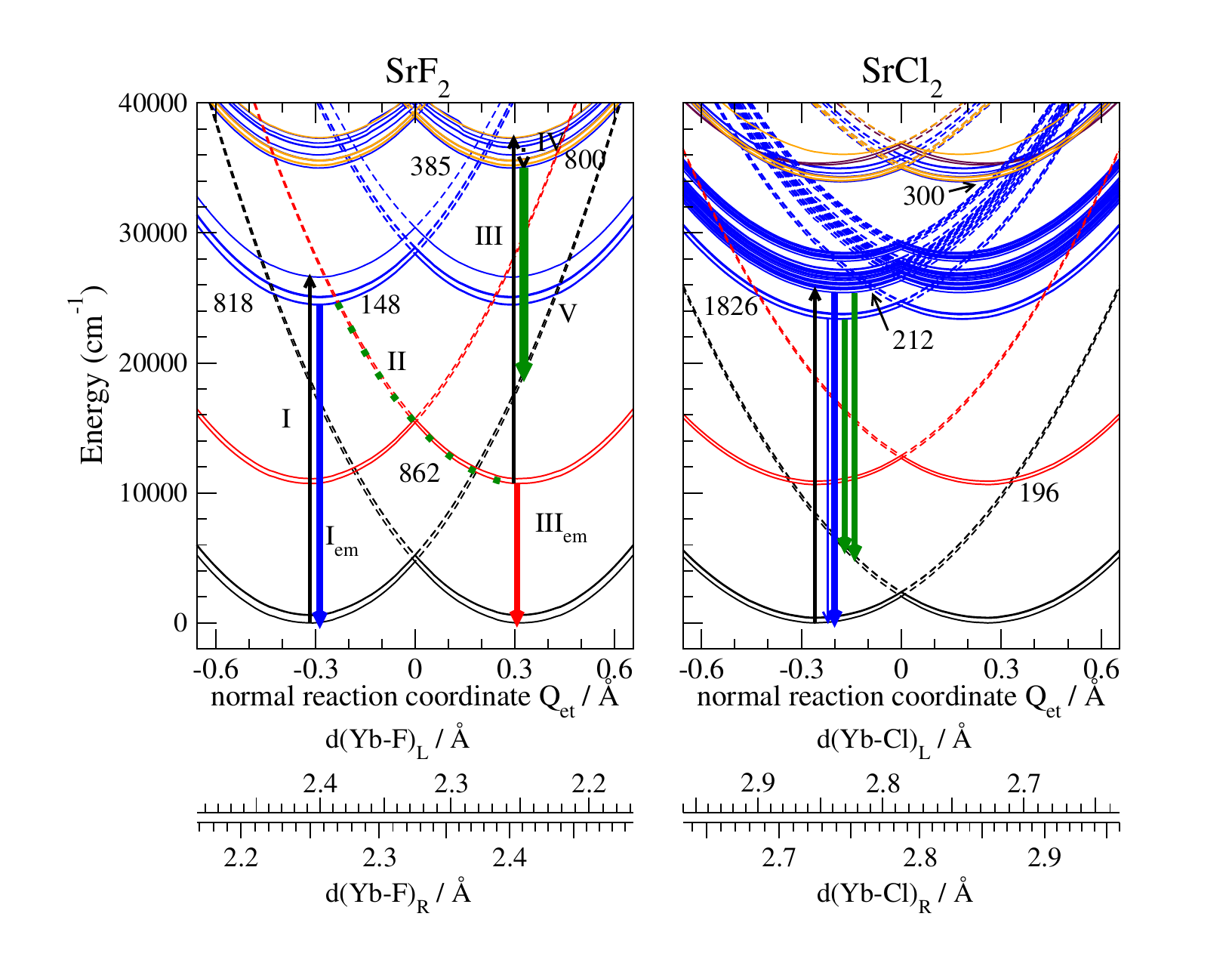}
}
\caption{
         Results of quantum mechanical calculations of the diabatic IVCT energy diagram for
         \Ybiipiiip\ pairs in Yb-doped  SrF$_2$, and SrCl$_2$ 
         crystals along the ground state normal
         electron transfer reaction coordinate Q$_{et}$.
         IVCT luminescence in SrF$_2$:
         steps I to V;
         excitation of the IR luminescence of \Ybthp: steps I, II, III$_{em}$;
         regular $5d-4f$ emission: steps I, I$_{em}$.
         SrCl$_2$:
         absorption (black), $5d-4f$ emissions (blue) and IVCT emissions (green)  
         are indicated by arrows.
         Energy barriers in \cmm1\ are indicated next to the crossing points between
         two electronic states of the pairs; see details in Table~\ref{TAB:IVCTbarr}.
         See caption of Fig.~\ref{FIG:CaF2-monomers-dimer} and text for details.
        }
\label{FIG:SrF2Cl2-YbIIYbIII-11-few}
    \end{figure}
    \def\escalafig{0.7}\clearpage
    \begin{figure}[ht]
\resizebox{\escalafig\textwidth}{!}{
  \includegraphics{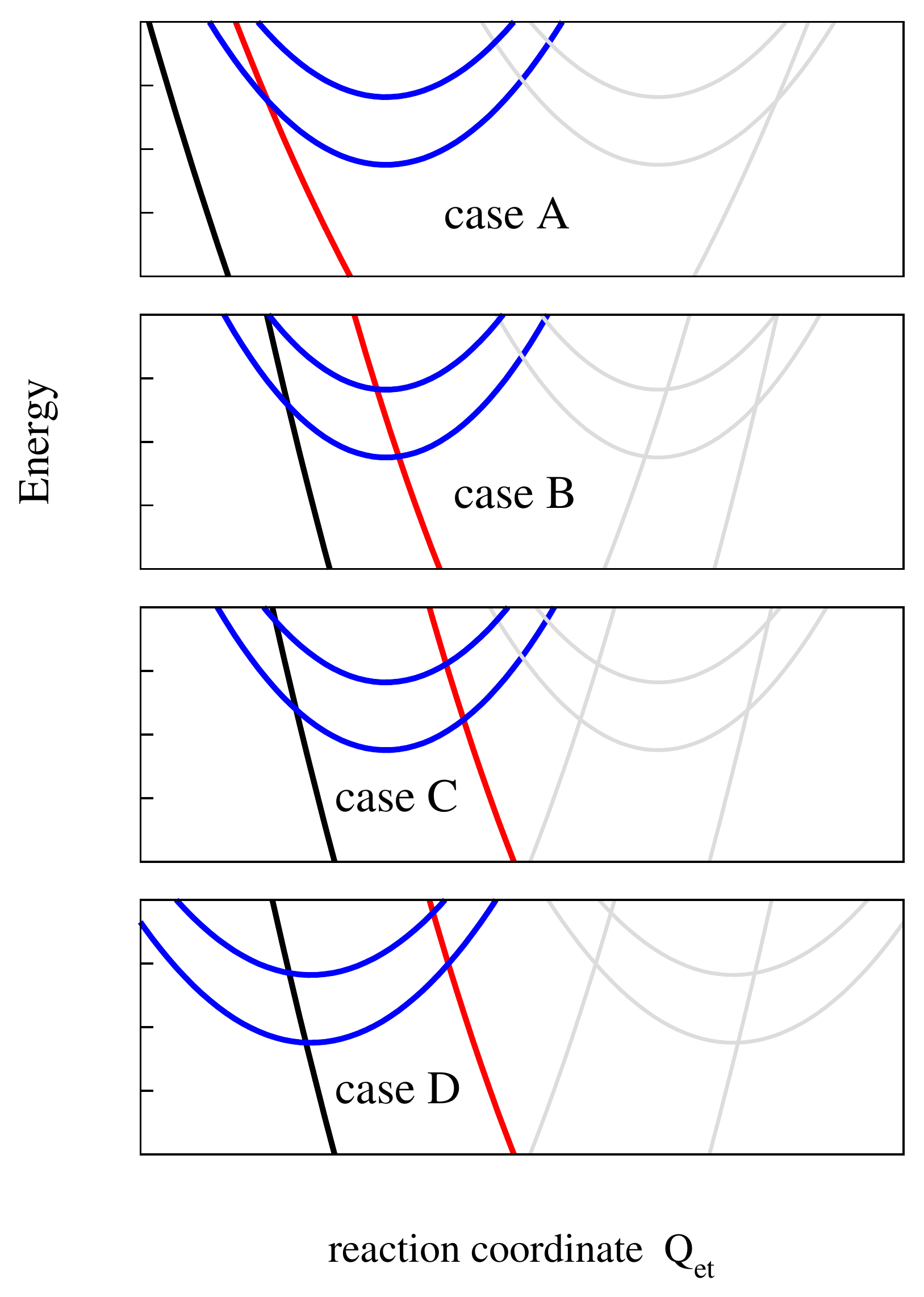}
}
\caption{
         Model cases for non-radiative electron transfer decays from \Ybiipiiip\ [\dFsh$5de_g$,\dFsh]
         excited states (blue) in Yb-doped fluorite crystals determining the luminescence properties.
         Two non-radiative decay pathways leading to IVCT luminescence excitation (red) or to
         luminescence quenching (black) are emphasized.
         \Ybtp-doped CaF$_2$ corresponds to case B,
         SrF$_2$ to a case between B and C,
         BaF$_2$, to case D, and SrCl$_2$ to case A.
        }
\label{FIG:Yb-cases}
    \end{figure}
}{}

\iftoggle{usualpreprint}{
    \clearpage \section*{Figure captions}
    \begin{figure}[ht]
      \input{FIG-\FIGa-cap.tex}
    \end{figure}
    \begin{figure}[ht]
      \input{FIG-\FIGb-cap.tex}
    \end{figure}
    \begin{figure}[ht]
      \input{FIG-\FIGc-cap.tex}
    \end{figure}
    \begin{figure}[ht]
      \input{FIG-\FIGd-cap.tex}
    \end{figure}
    \begin{figure}[ht]
      \input{FIG-\FIGe-cap.tex}
    \end{figure}
    \begin{figure}[ht]
      \input{FIG-\FIGf-cap.tex}
    \end{figure}
    \begin{figure}[ht]
      \input{FIG-\FIGg-cap.tex}
    \end{figure}
    \begin{figure}[ht]
      \input{FIG-\FIGh-cap.tex}
    \end{figure}
    \begin{figure}[ht]
      \input{FIG-\FIGi-cap.tex}
    \end{figure}
    \begin{figure}[ht]
      \input{FIG-\FIGj-cap.tex}
    \end{figure}
    \vfill\mbox{}
    \def\escalafig{0.9}
    \clearpage\begin{center}
    \input{FIG-\FIGa-fig.tex}
    \vfill
    Figure~\ref{FIG:\FIGa}. 
    \end{center}\thispagestyle{empty}
    \def\escalafig{0.9}
    \clearpage\begin{center}
    \input{FIG-\FIGb-fig.tex}
    \vfill 
    Figure~\ref{FIG:\FIGII}. 
    \end{center}\thispagestyle{empty}
    \def\escalafig{0.9}
    \clearpage\begin{center}
    \input{FIG-\FIGc-fig.tex}
    \vfill 
    Figure~\ref{FIG:\FIGIII}. 
    \end{center}\thispagestyle{empty}
    \def\escalafig{0.9}
    \clearpage\begin{center}
    \input{FIG-\FIGd-fig.tex}
    \vfill 
    Figure~\ref{FIG:\FIGIV}. 
    \end{center}\thispagestyle{empty}
    \def\escalafig{0.9}
    \clearpage\begin{center}
    \input{FIG-\FIGe-fig.tex}
    \vfill 
    Figure~\ref{FIG:\FIGV}. 
    \end{center}\thispagestyle{empty}
    \def\escalafig{0.9}
    \clearpage\begin{center}
    \input{FIG-\FIGf-fig.tex}
    \vfill 
    Figure~\ref{FIG:\FIGVI}. 
    \end{center}\thispagestyle{empty}
    \def\escalafig{0.9}
    \clearpage\begin{center}
    \input{FIG-\FIGg-fig.tex}
    \vfill 
    Figure~\ref{FIG:\FIGVII}. 
    \end{center}\thispagestyle{empty}
    \def\escalafig{0.9}
    \clearpage\begin{center}
    \input{FIG-\FIGh-fig.tex}
    \vfill 
    Figure~\ref{FIG:\FIGVIII}. 
    \end{center}\thispagestyle{empty}
    \def\escalafig{0.9}
    \clearpage\begin{center}
    \input{FIG-\FIGi-fig.tex}
    \vfill 
    Figure~\ref{FIG:\FIGIX}. 
    \end{center}\thispagestyle{empty}
    \def\escalafig{0.9}
    \clearpage\begin{center}
    \input{FIG-\FIGj-fig.tex}
    \vfill 
    Figure~\ref{FIG:\FIGX}. 
    \end{center}\thispagestyle{empty}
}{}


\clearpage 
\end{widetext}
\end{document}